\documentclass[11pt]{article}
\usepackage[a4paper,margin=1in]{geometry}
\usepackage{fontspec}
\usepackage{amsmath,amssymb}
\usepackage{graphicx}
\usepackage{booktabs,longtable,array,ragged2e,pdflscape}
\usepackage{caption}
\usepackage{microtype}
\usepackage{xurl}
\usepackage{hyperref}
\usepackage{enumitem}
\hypersetup{colorlinks=true,linkcolor=black,citecolor=black,urlcolor=blue}
\graphicspath{{./}}
\title{Operationalizing open-ended biological discovery across single-cell representations}
\author{}
\date{}
\begin{document}
\maketitle

\textbf{Authors:} Ningxuan Zhang\textsuperscript{1}, Ziwei Wang\textsuperscript{2}, Ning Xie\textsuperscript{2}, Na Liu\textsuperscript{3,*}

\textbf{Affiliations:} \textsuperscript{1}Department of Biochemistry, University of Oxford, South Parks Road, Oxford OX1 3QU, United Kingdom; \textsuperscript{2}Department of Gastroenterology, The Second Affiliated Hospital of Xi'an Jiaotong University, Xi'an 710004, P.R. China; \textsuperscript{3}Department of Gastroenterology, Hainan General Hospital (Hainan Affiliated Hospital of Hainan Medical University), Haikou 570311, P.R. China

\textbf{Correspondence:} Na Liu (liunafmmu@hainmc.edu.cn)

\begin{abstract}

Single-cell studies are typically initiated from predefined research questions, leaving much of the biological information encoded within existing data unexplored. We formalize open-ended discovery as an analytical paradigm, in which data-derived signals are identified before biological context is interrogated and subsequently evaluated according to their potential to justify prospective experimental investment. Here we develop PROSPECTor, an end-to-end framework that searches for reproducible biological structures across conventional expression representations and diverse foundation-model embeddings, translating robust signals into quantitatively testable candidate hypotheses. Projection into unseen datasets then evaluates their generalizability and phenotype association, providing a scalable screen for candidates that warrant prospective validation. Supported signals emerged from different representation spaces and search strategies. PROSPECTor-nominated hypotheses were then examined in independent biological settings: fibroblast extracellular-matrix programmes demonstrated transferability to an independent mouse cohort with an intervention context, while a patient-resolved gastric-cancer T-cell programme recurred across single-cell, bulk and spatial cohorts. PROSPECTor establishes an auditable framework for systematically revisiting single-cell datasets across expanding representation spaces, turning retrospective collections into prospective resources for biological discovery that can motivate new research questions.

\end{abstract}

\section*{Introduction}

Single-cell RNA sequencing (scRNA-seq) resolves biological variation across multiple scales, from shifts in cellular composition and discrete cell states to continuous functional axes and coordinated transcriptional programmes.\textsuperscript{1,2} Recent autonomous systems further suggest that substantially more signals can be extracted from existing single-cell datasets than is typically reported.\textsuperscript{3} Yet most analyses begin once an investigator has specified a biological question of interest, whether a disease, intervention, mechanism or cell population, and selected an analysis strategy suited to that focus\textsuperscript{1,2,4} (Fig. 1a). This question-driven paradigm is useful for testing defined hypotheses, but samples only a researcher-selected subset of the structures potentially present in the high-dimensional data, while failed or deprioritized exploratory branches are rarely preserved.\textsuperscript{4,5} However, whether existing single-cell datasets can serve as search spaces for discovering additional, generalizable biological hypotheses remains largely unexplored.

Pretrained single-cell foundation models (scFMs) expand this opportunity by providing additional observation spaces over the same transcriptomic data. Learned cell representations encode molecular variation according to model-specific objectives and training procedures, creating alternative organizations of cellular information beyond conventional expression spaces.\textsuperscript{6–9} These latent spaces may emphasize heterogeneous aspects of cellular biology, such that cell states, transcriptional programmes and condition-associated signals that are prominent in one representation may be less apparent in another.\textsuperscript{7–10} This motivates asking whether scFM-derived representations can be treated not only as tools for predictive downstream tasks, but also as complementary discovery substrates that can be searched alongside conventional expression-based spaces such as highly variable gene principal-component analysis (HVG–PCA).

Recent scientific agents increasingly automate literature retrieval, hypothesis generation, tool selection, code execution and biological data analysis.\textsuperscript{3,11–14} Yet these systems typically operate from predefined goals, research context or source-study information.\textsuperscript{3,11–13} An alternative and less studied mode would begin with a context- and phenotype-blind search for patterns that emerge from the data, separating data-first discovery from workflows in which hypothesis generation, knowledge retrieval and analysis are interleaved.\textsuperscript{14}

This reframes the task from addressing a predefined biological question to independently identifying signals that can motivate the next research question. Such open-ended discovery is intrinsically difficult to evaluate, as the relevant findings are not known in advance. Existing evaluations often ask whether generated analyses reproduce published conclusions or rely on expert judgements of plausibility or scientific soundness.\textsuperscript{3,11,12} The former is poorly suited to open-ended discovery, since genuinely novel signals have no published counterpart against which they can be evaluated. Prospective experiments can establish whether such signals reflect genuine biology,\textsuperscript{10–12} but cannot feasibly validate every candidate produced by large-scale search. Valid single-cell inference requires biological replication rather than treating individual cells as independent observations.\textsuperscript{15,16} Cross-study evaluation must also contend with technical variation and differences in sample composition.\textsuperscript{16–18} We therefore designed a scalable intermediate evaluation that tests transferability across independent donors and cohorts, robustness to technical and compositional covariates, and generalizability of phenotype association.

Here we introduce PROSPECTor, a system that formalizes open-ended discovery from single-cell representations as an auditable and systematic process of signal search and evidence acquisition (Fig. 1b). In addition to the conventional bioinformatics observation spaces in which existing autonomous single-cell workflows largely operate, PROSPECTor receives diverse single-cell representations and searches for structures that recur across independent biological units. PROSPECTor then traces candidate structures back to the cells in which they arise and the molecular programmes they reflect, tests plausible technical and compositional explanations, and finally fixes candidates as quantitatively defined hypotheses for evaluation in external data. We use hypothesis generation in an operational, pre-mechanistic sense: the construction of a falsifiable candidate that specifies a cellular context, a measurable gene programme, a putative functional interpretation and a rejection criterion in unseen data.

We pair PROSPECTor with a prospective-style evaluation framework that assesses candidate hypotheses across diverse settings alongside reference workflows, and systematically examine how six pretrained scFMs reorganize single-cell measurements and reshape the landscape for biological discovery. Finally, we follow PROSPECTor-nominated hypotheses, including two fibroblast extracellular-matrix programmes and a patient-resolved gastric T-cell programme evaluated across single-cell, bulk and spatial data. Across these analyses, PROSPECTor complements conventional question-driven research with a signal-first approach: reproducible signals are identified before their biological context is interrogated, and subsequent evidence is used to prioritize candidates for further investigation. Together, this study provides a framework for revisiting existing data across an expanding space of representations, recasting these data as prospective resources from which new biological questions can emerge.

\section*{Results}

\subsection*{Operationalizing discovery as a search-and-evidence-acquisition process}

PROSPECTor operationalizes open-ended discovery by identifying biological signals in a dataset before a biological question is specified or phenotype labels are revealed (Fig. 1b). During discovery, the system receives molecular measurements, gene identifiers and anonymized identifiers for cells, samples, donors and datasets. PROSPECTor searches multiple representation spaces derived from the same single-cell data: HVG–PCA provides a conventional expression representation, while embeddings from the pretrained scFMs, including scGPT, Geneformer, scFoundation, UCE, Tahoe-X1 and STATE, provide six additional representations of cellular variation.\textsuperscript{6,19–23} Each representation is searched independently in its native dimensions and in a standardized 32-dimensional projection, preserving model-specific structure while providing a matched low-dimensional analysis.

Since potentially meaningful biological signals from single-cell data may occupy several geometrical forms, PROSPECTor applies four complementary unsupervised detectors to each observation space (Fig. 1c). Stable clustering identifies discrete populations that recur across cell resampling, while sparse-direction analysis and rare-state detection identify reproducible continuous gradients or rare cell neighbourhoods represented across donors. In parallel, expression-programme analysis uses cNMF to recover coordinated gene modules directly from measured counts.\textsuperscript{24} Each detector produces a numerical candidate object that can be localized, decoded and projected.

Candidate generation initiates a structured evidence-acquisition process in PROSPECTor (Fig. 1d). Each candidate enters a directed graph in which successive analyses evaluate donor recurrence, cellular localization, molecular composition and sensitivity to measured technical or compositional explanations. Each branch supports a progressively more specific claim and must satisfy its own evidence requirements. Unsupported routes are closed, whereas candidates without an applicable measurement remain unresolved. The graph retains positive, negative and unavailable results so that the complete trajectory of candidate evaluation can be reconstructed.

A finalized candidate hypothesis contains a fixed quantitative definition, a signed gene programme, a cellular scope when localization is supported, a putative functional interpretation derived from molecular analysis and a prespecified rejection criterion for held-out evaluation. The same hypothesis can then be evaluated for recurrence, cellular localization and phenotype association without feature reselection, coefficient fitting, threshold optimization or direction reversal. The complete graph records how each hypothesis was constructed, which observations support it and where its interpretation remains limited. Together with the evaluation procedure, this design makes the data-driven, open-ended discovery process systematic and auditable.

\clearpage
\begin{center}
\includegraphics[width=\textwidth,height=0.72\textheight,keepaspectratio]{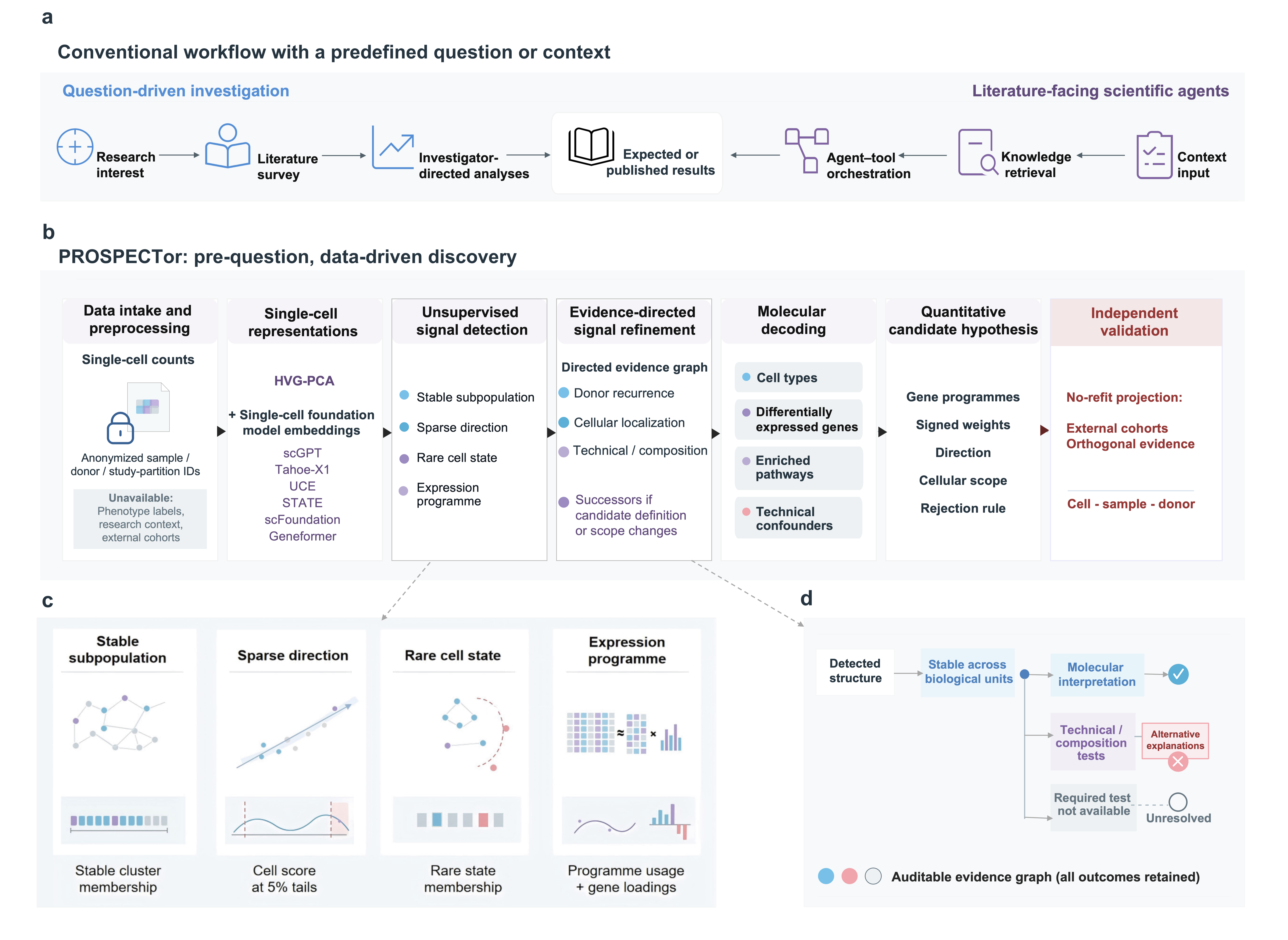}
\end{center}
\begingroup\small
\noindent\textbf{Figure 1 | Data-driven signal discovery with PROSPECTor.} \textbf{a,} Conventional question-led single-cell analysis (left) and literature-facing scientific-agent evaluation (right), illustrating how researcher choices, prior context and answer retrieval can constrain the search. \textbf{b,} PROSPECTor workflow from raw-data intake and construction of different single-cell representations through unsupervised signal detection, evidence acquisition, molecular interpretation, candidate fixation and independent-cohort evaluation. \textbf{c,} Four signal-detection operators and their numerical outputs, comprising stable populations, sparse directions, rare states and expression programmes. \textbf{d,} Evidence acquisition through a directed graph in which successive analyses assess donor recurrence, cellular localization, molecular evidence and technical or compositional explanations. The terminal state of a candidate hypothesis is frozen, rejected or unresolved.\par
\endgroup

\subsection*{Prospective-style evaluation across unseen cohorts and discovery strategies}

Since open-ended discovery has no predefined target, we evaluated whether data-derived hypotheses could transfer to unseen cohorts and retain a phenotype-associated effect at the biological-unit level. Each discovery workflow was evaluated in 22 runs (Fig. 2a). The primary analysis comprised 15 leave-one-study-out runs across COVID-19 blood,\textsuperscript{25–31} idiopathic pulmonary fibrosis (IPF),\textsuperscript{32–35} mouse colonic inflammation\textsuperscript{36–38} and paired gastric cancer.\textsuperscript{39–41} One additional COVID-19 fold containing only one control donor was retained as a sensitivity analysis, while six source-paper rediscovery runs were analysed separately (Supplementary Table 1).\textsuperscript{39–44} In each primary fold, one complete study and its phenotype labels remained unavailable until discovery was complete. Before these labels were revealed, each candidate's genes, signed weights, orientation, normalization, aggregation rule and rank were fixed. The unchanged programme was projected into held-out cells, then aggregated through samples to the donor, patient or animal as the unit of inference. Supported candidates@4 counted candidates among the frozen top four with an oriented Hedges \emph{g}\textsuperscript{45} of at least 0.5 in the held-out cohort. A candidate qualified only when this threshold was met for both the original programme and a second projection after removing genes overlapping the best-matching reference cell-marker set, using the marker libraries described in Methods. The two projections tested no-refit transfer and whether the effect persisted after removal of a familiar cell-identity programme.

CellVoyager and Biomni were included as reference workflows under the same phenotype-hidden discovery and frozen-projection design.\textsuperscript{3,46} At the primary threshold, the equally weighted mean across settings was 1.54 for PROSPECTor, 0.72 for Biomni and 0.58 for CellVoyager (Fig. 2b); threshold sensitivity is shown in Fig. 2c. Values varied across settings and were not treated as a pooled biological comparison. Source consistency, cellular localization, phenotype association within the nominated cellular context, technical robustness, composition robustness and six-study source-paper rediscovery were reported separately (Fig. 2d). Limited concordance with registered source-study findings across settings and across workflows highlighted the incompleteness of rediscovery as an evaluation target for open-ended search, where unreported signals and failed trajectories are unavailable as reference outcomes.

Search breadth altered the candidates recovered across the four IPF holdouts. Omitting any of the four detection strategies removed candidates not recovered by the remaining operators (Fig. 2e). Searches across all seven representations produced 64 final candidates, compared with 16 using HVG–PCA alone; ten and five, respectively, showed phenotype association at \emph{q} ≤ 0.05 after held-out projection (Fig. 2f). Native and standardized 32-dimensional searches also recovered partially different candidate sets. Both the search operator and the representation therefore affected which hypotheses reached evaluation.

Progressively truncating the recorded analysis trajectories recovered fewer candidates that later showed phenotype association, indicating that the deterministic evidence-acquisition graph accumulated information useful for candidate retention beyond initial signal detection (Supplementary Fig. 1a,b). By contrast, outcome-hidden language-model reranking substantially reordered the pre-validation top four but left the number of phenotype-associated candidates unchanged in F01–F03 and increased it by only one in F04, providing little evidence of consistent gain (Supplementary Fig. 1c). We therefore kept signal discovery and evidence acquisition deterministic: allowing semantic reasoning to influence signal discovery would make the information boundary required for question-unspecified discovery harder to audit. Annotation relabelling produced a mean recovery of 0.90. Cell resampling yielded a median membership Jaccard of 0.865 for discrete candidates and a median score correlation of 0.983 for continuous candidates. Changing the HVG set from 2,000 to 3,000 genes yielded a median overlap of 0.122 (Supplementary Fig. 1d). This low overlap indicates that candidate recovery in the evaluable subset was sensitive to the precise HVG definition, despite greater stability under cell resampling.

\clearpage
\begin{center}
\includegraphics[width=\textwidth,height=0.66\textheight,keepaspectratio]{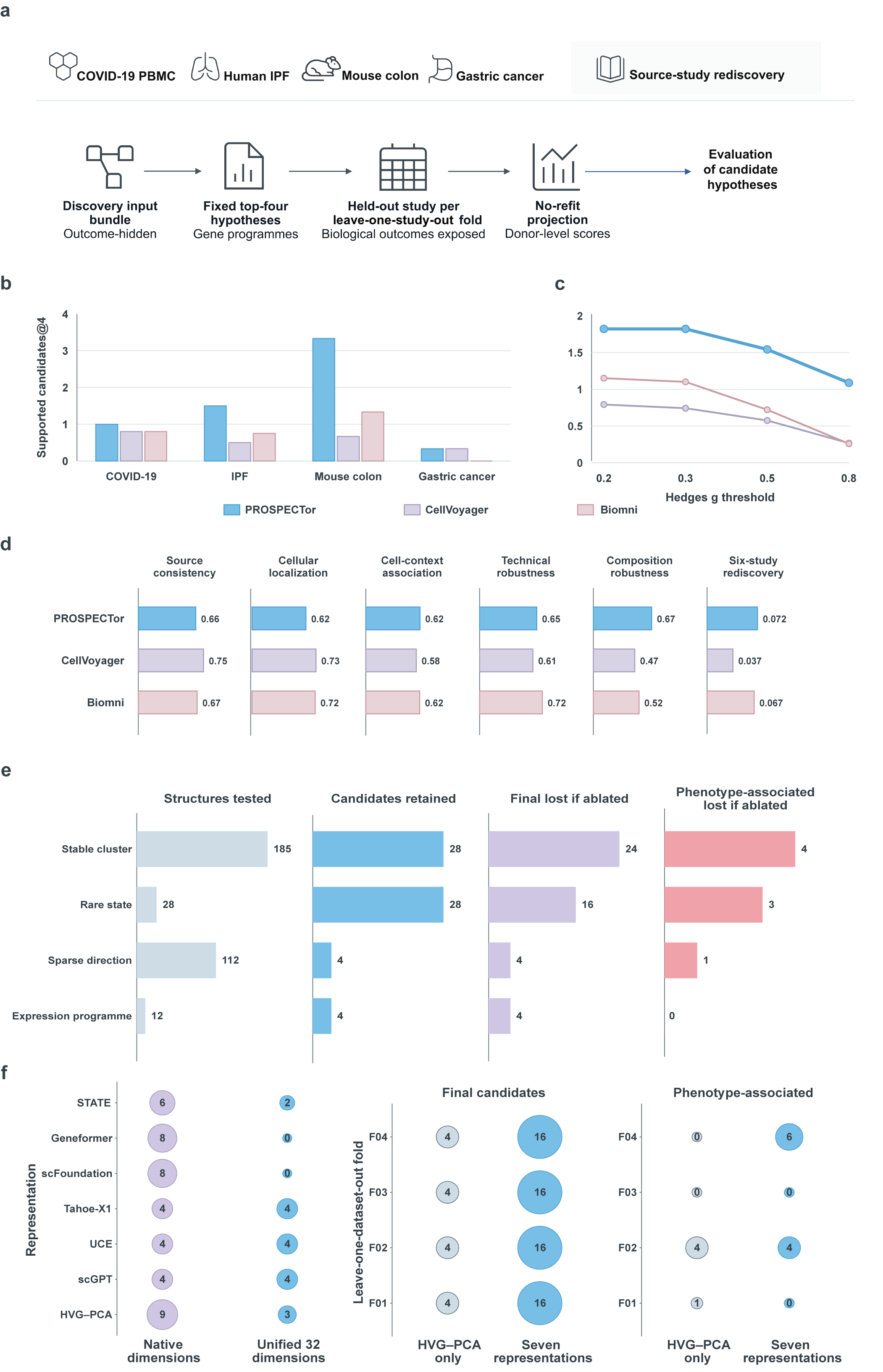}
\end{center}
\begingroup\small
\noindent\textbf{Figure 2 | Prospective-style evaluation of candidate hypotheses and PROSPECTor components.} \textbf{a,} Four study-level holdout settings and a separate six-study source-study rediscovery setting. In each holdout fold, one complete study remains unavailable until candidate hypotheses have been fixed. The frozen hypotheses are then projected into the held-out study without refitting, and candidate support is evaluated at the biological-unit level. \textbf{b,} Mean Supported candidates@4 in COVID-19 blood, human IPF, mouse colon and gastric cancer for PROSPECTor and the two reference workflows, CellVoyager and Biomni. Fold-level counts were averaged within COVID-19 (five primary folds), human IPF (four), mouse colon (three) and gastric cancer (three); bars show setting means. \textbf{c,} Equally weighted means across the four settings at Hedges \emph{g} thresholds of 0.2, 0.3, 0.5 and 0.8. Lines connect the same discovery workflow across thresholds. \textbf{d,} Source consistency, cellular localization, phenotype association within the declared cell population, technical robustness, composition robustness and deterministic source-paper rediscovery. The five evidence measurements were averaged from candidates to folds and then to evaluation settings; rediscovery was averaged across registered findings and source studies. \textbf{e,} Search output and operator analysis across IPF folds F01–F04. For each operator, the panel shows the number of structures tested, candidate signals retained, final candidates lost when that operator and its descendants were removed, and independently phenotype-associated candidates lost in the same analysis. \textbf{f,} Left, candidate signals detected from each representation in its native dimensionality or after projection to a common 32-dimensional space. Centre and right, final candidates and candidates with independent phenotype association from matched searches using all seven representations or HVG-PCA alone.\par
\endgroup

\subsection*{Foundation-model representations expose heterogeneous spaces for discovery}

We examined how different scFM-derived representations contributed to discovery by focusing on IPF F04, the fold with the largest number of phenotype-associated candidates originating from scFM spaces (Fig. 2f). The same 10,000 cells sampled from F04 were represented in HVG–PCA and six extracted scFM embeddings (Fig. 3a). Candidate detection used the corresponding high-dimensional representations, while UMAP was used only for visualization here.\textsuperscript{47} The same cells occupied markedly different local arrangements across representations. Some spaces strongly separated broad cell identities, whereas others emphasized variation within lineages or placed multiple cellular contexts within shared local neighbourhoods.

We quantified these differences using four measurements of complementary properties of each representation (Fig. 3b). Cross-donor identity measured whether a cell's nearest neighbour from another donor shared the same broad cell-type annotation. Cross-study disease signal assessed whether donor-level cell-type aggregates retained information associated with disease status. Donor mixing measured the fraction of local neighbours contributed by other donors, and technical encoding assessed how well each representation predicted detected-gene counts under donor-grouped cross-validation. Most representations strongly preserved broad cell identity across donors. UCE showed lower single-cell identity agreement (0.40), stronger donor mixing (0.96), weaker cross-study disease association (AUROC 0.58) and limited depth prediction (R\textsuperscript{2} = 0.20). The complete values and resampling intervals are shown in Fig. 3b. These measurements varied independently across representations, with no single measurement accounting for all observed differences.

We next compared local geometric consistency directly using the overlap between 30-nearest-neighbour sets for the same cells in different representations (Fig. 3c and Supplementary Fig. 2d). Tahoe-X1 and scFoundation showed the greatest overlap among distinct representations, with a mean Jaccard index of 0.46. By contrast, the mean neighbourhood overlap between UCE and every other representation was below 0.004. This divergence was less pronounced after aggregation. At the donor and broad cell-type level, UCE recovered broad identity with an agreement of 0.935 and retained a cross-study disease AUROC of 0.700. Local neighbourhood agreement and preservation of broader cellular structure therefore differed substantially even for the same representation.

The distinctive organization of cells in UCE motivated closer examination of a UCE-derived rare-state candidate from the same F04 run (Fig. 3d). The unsupervised detector assigned every cell a continuous rarity score and fixed the upper 5\% before case or control labels were available. Subsequent cell-context and differential-expression analyses decoded the signal into a fixed gene programme. Projection of this unchanged programme into the 60-donor GSE136831 study\textsuperscript{35} yielded an AUROC of 0.83 and a Benjamini–Hochberg-adjusted \emph{q} = 0.008.\textsuperscript{48} Thus, a representation with little local-neighbourhood agreement with conventional expression or other scFM embeddings nevertheless generated a quantitatively defined programme that retained phenotype association in an independent cohort.

Across all four IPF folds, we further examined whether scFM-derived candidates recurred in another representation, retained information beyond conventional HVG–PCA and showed phenotype association after independent projection (Fig. 3e; Supplementary Fig. 2b,c). Among 32 final scFM-derived candidates, 14 recurred in another representation. Five of the six candidates with independent phenotype association did not recur in another representation. Cross-representation recurrence and independent phenotype association therefore captured different properties of the candidate signals.

Together, these observations suggest an additional use of pretrained single-cell models beyond downstream prediction. Pretraining produces distinct latent representations of the same molecular measurements, and different models can organize cellular variation in ways that emphasize different structures. PROSPECTor uses these embeddings as complementary substrates for unsupervised search and systematically evaluates the biological signals identified in each representation space.

\clearpage
\begin{center}
\includegraphics[width=\textwidth,height=0.72\textheight,keepaspectratio]{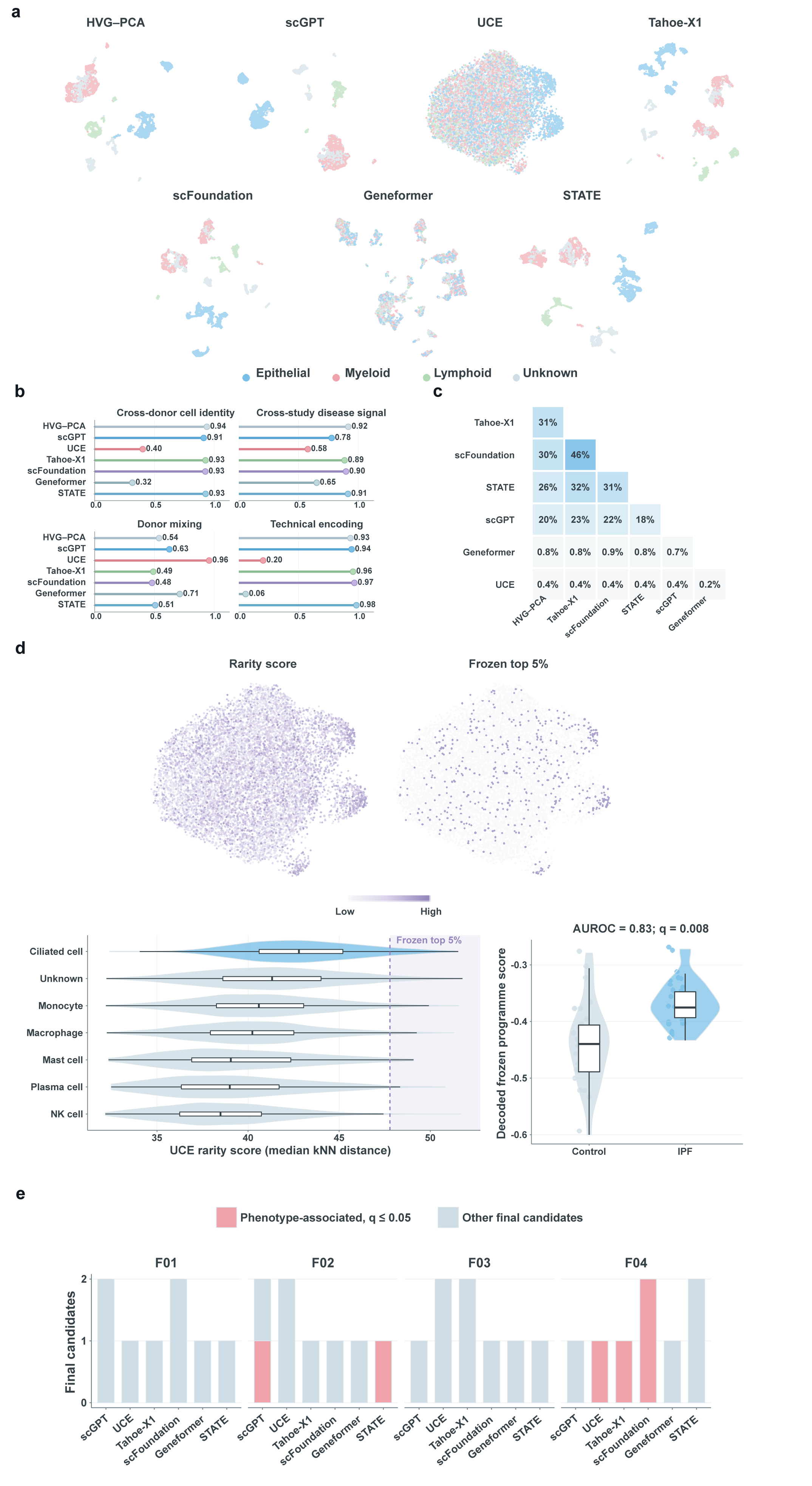}
\end{center}
\begingroup\small
\noindent\textbf{Figure 3 | Single-cell representations expose distinct biological structures and supported signals.} \textbf{a,} UMAPs shown for visualization for the same 10,000 F04 cells in HVG–PCA and six scFM representations. Colour denotes broad cell identity. \textbf{b,} Cross-donor cell-identity agreement, cross-study disease classification, donor mixing and technical encoding for each representation. Points show full-data estimates. For cross-donor identity and cross-study disease signal, horizontal intervals show the 2.5th–97.5th percentiles across 20 donor-stratified 80\% resamples; donor mixing and technical encoding are shown without resampling intervals. Each measurement is shown on its own scale. \textbf{c,} Pairwise Jaccard overlap between 30-nearest-neighbour sets for the same cells across different representations. \textbf{d,} UCE rare-state example, showing the identification and evaluation of this candidate signal: the high-dimensional rarity score, fixed upper-5\% membership, rarity-score distributions across cell identities and no-refit projection of its decoded gene programme into the held-out 60-donor GSE136831 study. This independent projection achieved an AUROC of 0.83 and \emph{q} = 0.008. \textbf{e,} Final candidates contributed by each representation and fold, separated according to whether they showed independent phenotype association at \emph{q} ≤ 0.05.\par
\endgroup

\subsection*{Nominated fibroblast hypotheses connect rediscovery, transferability and an intervention context}

To assess whether PROSPECTor-generated hypotheses for mouse colonic inflammation generalized beyond the discovery datasets, we evaluated the leading candidate hypotheses in an independent mouse cohort previously generated within the study team, which included an additional intervention context with BAPN treatment (Fig. 4a).\textsuperscript{38}

The discovery analysis integrated two published single-cell datasets of mouse DSS colitis, containing 16 mice and 103,686 cells (GSE264408 and GSE172261).\textsuperscript{36,37} Six hypotheses were fixed before validation in the external cohort. Candidate hypothesis \#1 originated from a recurrent cell subpopulation, whereas candidate \#2 originated from a continuous expression direction (Fig. 4b), and both structures were primarily localized to fibroblasts. Molecular decoding associated candidate \#1 with Pcolce2, Rbp1 and Serpinf1, and candidate \#2 with Tgfbi, Igf2 and Bmp3. The two independently derived programmes converged on fibroblast activation and extracellular-matrix remodelling processes reported in the GSE172261 source study during inflammatory colon injury.\textsuperscript{37} This shared interpretation was consistent with broader evidence for fibroblast and mesenchymal remodelling in inflammatory bowel disease and colitis-associated cancer.\textsuperscript{49,50} PROSPECTor therefore linked source-study rediscovery with quantitative hypothesis generation by defining two programmes that would subsequently be tested in independent mouse models.

All six candidate hypotheses were thus projected, with unchanged genes, weights and directions, into our validation dataset, generating fixed candidate scores for each mouse. The primary disease comparison included three Control mice and three AOM–DSS mice. Candidate \#2 showed the largest standardized disease-associated difference in the computed candidate score among the six projected hypotheses (Hedges \emph{g} = −2.56, approximate 95\% confidence interval −4.95 to −0.17), while candidate \#1 showed a concordant but less precise effect (\emph{g} = −1.29, −3.12 to 0.54; Fig. 4c). The remaining candidates showed weaker or inconsistent changes, indicating that transferability into an independent cohort was selective. Given the small cohort size, these effect estimates were interpreted descriptively.

We next asked whether the candidate effects remained apparent within the fibroblast population implicated during discovery. Fibroblasts were identified in the same six validation mice, and the candidate scores were averaged only within these cells. The fibroblast-restricted score of candidate \#1 decreased from 0.710 in Control to 0.454 in AOM–DSS mice, whereas that of candidate \#2 decreased from −0.305 to −0.420 (Fig. 4d), indicating that both transferred signals were measurable within fibroblasts in the external cohort.

Three additional AOM–DSS mice from the independent cohort received BAPN treatment, providing a separate intervention context beyond the case–control analysis. To compare treatment-associated shifts across candidate hypotheses, each readout was normalized to the observed AOM–DSS versus Control candidate-score difference, where zero represents the candidate-score mean in the AOM–DSS group and one represents the Control mean. The same calculation was applied to all available PROSPECTor and reference-workflow candidates (Fig. 4e), with candidate numbers referring to within-workflow ranks assigned during the evaluation run. The whole-sample candidate \#1 score exceeded the Control reference after BAPN treatment (normalized score movement = 1.996), whereas the fibroblast-restricted candidate \#1 score moved only 0.135 of the Control–AOM–DSS reference distance. Fibroblast abundance also shifted towards the Control reference, while the fibroblast-restricted candidate \#2 score moved further away from the Control reference (normalized movement = −0.390; Fig. 4f). These BAPN-associated shifts provide an exploratory view of gene programme modulation under matrix perturbation and do not establish rescue or treatment efficacy.

Taken together, this case shows how a signal identified without access to the original biological question can acquire biological context through subsequent evidence and become a candidate for prospective follow-up. These observations nominate a more specific question for wet-lab testing, for which multiplex fluorescence staining and subsequent mechanistic studies could provide experimental evidence.

\clearpage
\begin{center}
\includegraphics[width=\textwidth,height=0.63\textheight,keepaspectratio]{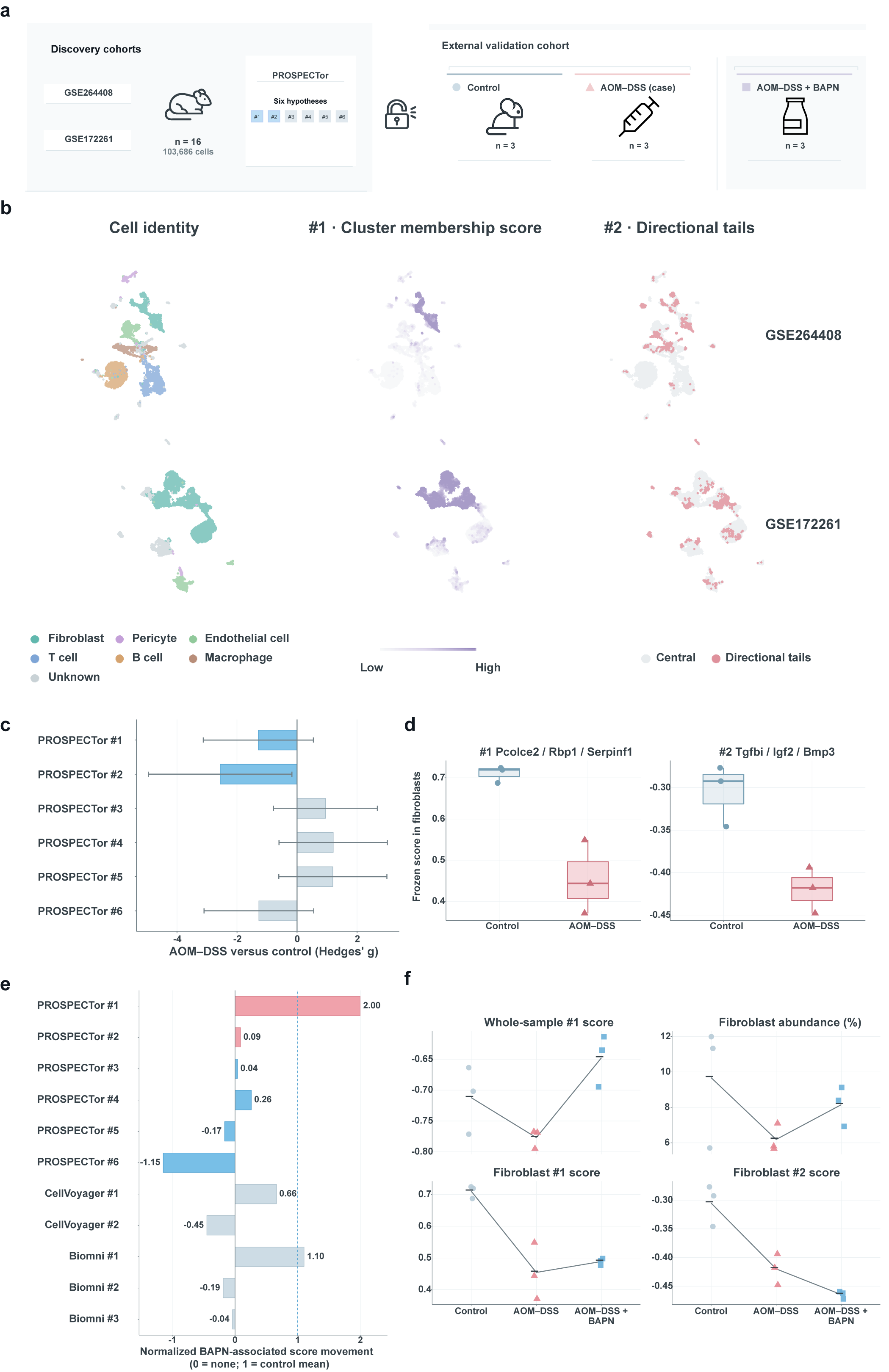}
\end{center}
\begingroup\small
\noindent\textbf{Figure 4 | Candidate hypotheses show cross-cohort transferability and differential associations under an intervention context.} \textbf{a,} Study design. Six candidate hypotheses were defined from 103,686 cells collected from 16 mice across GSE264408 and GSE172261 before access to the unseen validation cohort. The primary external disease comparison included three Control and three AOM–DSS mice. Three additional AOM–DSS mice receiving BAPN provided an exploratory matrix-perturbation arm. \textbf{b,} Descriptive visualization of the two leading fibroblast-associated candidates in the discovery studies. The left column shows broad cell identity in each two-dimensional embedding. The centre column shows the candidate \#1 detector metric, and the right column marks cells in the fixed tails of candidate \#2. \textbf{c,} Independent AOM–DSS versus Control effects for all six fixed candidates. Bars show mouse-level Hedges \emph{g}, and error bars show approximate 95\% confidence intervals. Negative values indicate lower scores in AOM–DSS. With three mice per group, the intervals are interpreted descriptively. \textbf{d,} Candidate \#1 and \#2 scores within independently annotated fibroblasts from the same six validation mice. Points represent mice. Boxes show the median and interquartile range, and whiskers extend to 1.5 times the interquartile range. \textbf{e,} Normalized BAPN-associated score movement for every PROSPECTor and reference-workflow candidate. The plotted value is (AOM–DSS + BAPN mean − AOM–DSS mean)/(Control mean − AOM–DSS mean). Zero denotes the AOM–DSS mean, and the dashed line at one denotes the Control mean. Values below zero or above one extend beyond the corresponding reference mean. Candidate numbers denote the order reported within each workflow during the evaluation runs. \textbf{f,} Whole-sample candidate \#1 score, fibroblast abundance, fibroblast-restricted candidate \#1 score and fibroblast-restricted candidate \#2 score. Points represent individual mice, and horizontal ticks show group means. Candidate scores retain candidate-specific units and vertical scales, and comparisons are therefore made only within each subpanel.\par
\endgroup

\subsection*{A gastric T-cell programme recurs across cancer cohorts and modalities}

We next investigated whether a hypothesis generated through label-blind discovery in cancer data could recur across independent patients, cohorts and measurement modalities. The gastric case study originated from discovery in GSE183904 with a complete study-level holdout in GSE270680 (Fig. 5a).\textsuperscript{40,41} The prioritized candidate arose from T-cell measurements in eight patients in GSE183904, and molecular decoding of that candidate signal produced a fixed 24-gene programme with signed weights. The programme was oriented once so that higher candidate scores indicated a shift towards the \emph{PSMB8}, \emph{PARP9}, \emph{SAMD9L} and \emph{GBP1} antigen-processing and interferon-stress pole.

The prespecified held-out dataset provided a first external verification. Sixteen patients in GSE270680 had at least 30 annotated T cells in both tumour and adjacent tissue, and 15 showed higher tumour-side scores. The mean paired difference was 0.244 and the exact sign-flip \emph{P} value was 6.1 × 10\textsuperscript{-5} (Fig. 5a). We additionally projected the fixed programme into four single-cell cohorts (Fig. 5b).\textsuperscript{39} Thirteen of the 23 projected patients in GSE150290 showed the expected score increase in tumour, with a mean paired shift of 0.027, providing directionally consistent but weak additional evidence. GSE206785 comprised matched tumour and adjacent normal gastric tissue from treatment-naive patients.\textsuperscript{51} Sixteen patients met the minimum of 30 annotated T cells in each tissue; 13 showed the expected increase in tumour, and the mean paired difference was 0.243 (95\% bootstrap confidence interval 0.139–0.351; exact sign-flip \emph{P} = 5.2 × 10\textsuperscript{-4}). The positive direction was retained across normalization, CD4/CD8 weighting, patient-level covariate adjustment and leave-one-gene-out sensitivity analyses. Two smaller studies supplied four and three eligible pairs, respectively (GSE167297 and GSE234129).\textsuperscript{42,43} Score changes for all four patients in the first cohort and two of three in the second were concordant, with mean differences of 0.085 and 0.235.

We asked whether the programme's effect recurred outside single-cell measurements. Applying the same coefficients to paired TCGA-STAD bulk profiles produced a positive tumour-minus-adjacent shift in 27 of 33 patients, with a mean paired difference of 0.511 (exact sign-test \emph{P} = 3.24 × 10\textsuperscript{-4}; Fig. 5c).\textsuperscript{52} This result supports recurrence of the programme in whole-tissue expression yet does not localize the signal to T cells.

We therefore used spatial transcriptomics to examine its localization. Ten GSE251950 Visium sections from nine patients were divided into T-rich and other regions using six pan-T-cell genes that did not overlap the examined programme.\textsuperscript{53} The programme was higher in T-rich regions in all nine patients, with a mean patient-level difference of 0.266 (bootstrap 95\% confidence interval, 0.188–0.347; one-sided exact sign-flip \emph{P} = 0.002; Fig. 5d,e). After adjustment for the continuous pan-T-cell score, eight of nine patients retained a positive effect, with a mean adjusted difference of 0.071 (95\% confidence interval, 0.036–0.106; \emph{P} = 0.006). These results support localization of the programme to multicellular T-rich regions.

Finally, the interpretation of this gastric-cancer-associated programme and the limits of the current evidence were reviewed with investigators with expertise in gastric-cancer biology, shaping the next research question and how it could be tested prospectively.

\clearpage
\begin{center}
\includegraphics[width=\textwidth,height=0.72\textheight,keepaspectratio]{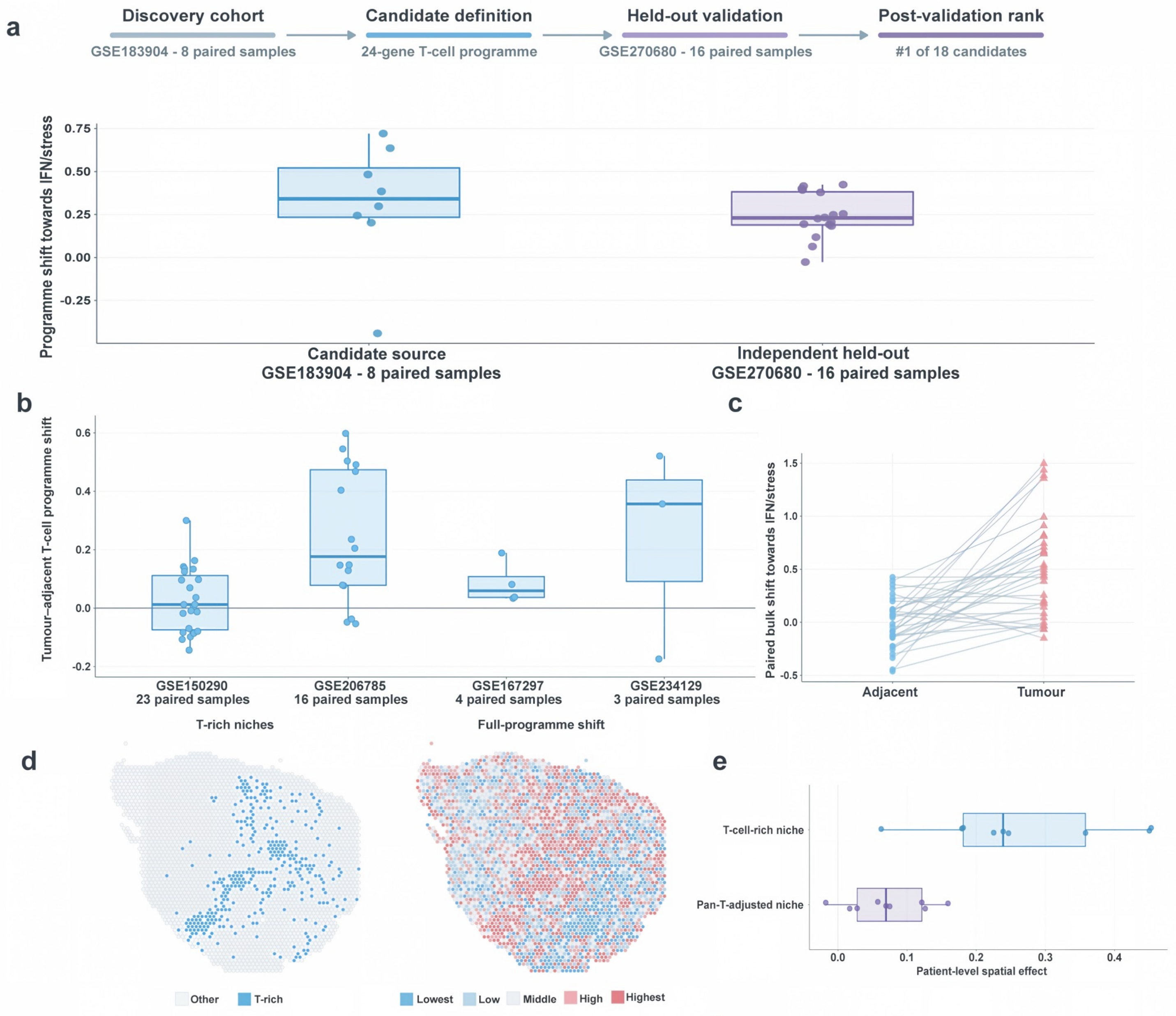}
\end{center}
\begingroup\small
\noindent\textbf{Figure 5 | Recurrence of a gastric-cancer-associated T-cell programme across cohorts and data modalities.} \textbf{a,} Discovery, held-out evaluation and post-validation ranking. The fixed gene programme was derived from eight patients with paired tumour and adjacent samples in GSE183904 before independent evaluation in GSE270680. Points show patient-level tumour-minus-adjacent shifts towards the interferon-stress pole for eight discovery patients and 16 validation patients. Boxes show the median and interquartile range. The programme ranked first among 18 candidates after the held-out evaluation. \textbf{b,} Projections in four single-cell cohorts: GSE150290 (\emph{n} = 23 patients), GSE206785 (\emph{n} = 16), GSE167297 (\emph{n} = 4) and GSE234129 (\emph{n} = 3). Each point represents one patient's tumour-minus-adjacent T-cell score difference. Positive values indicate a shift towards the antigen-processing and interferon-stress pole. Boxes show the median and interquartile range, and whiskers extend to 1.5 times the interquartile range. \textbf{c,} Orthogonal tissue-level validation in TCGA-STAD. The fixed gene coefficients were applied to bulk expression from 33 patients with paired adjacent and primary-tumour samples. Higher values indicate the interferon-stress direction, and lines connect samples from the same patient. \textbf{d,} Spatial localization in a GSE251950 Visium section selected for visualization by the greatest Jaccard overlap between independently defined T-rich spots and the upper candidate-score quintile. Left, T-rich spots defined using six pan-T-cell genes that do not overlap the candidate programme. Right, the transferred candidate score divided into within-section quantiles. Section selection did not affect the patient-level analyses. \textbf{e,} Patient-level spatial effects after combining sections from the same patient. The upper row shows score differences between T-rich and other regions. The lower row shows T-rich coefficients after adjustment for continuous pan-T-cell abundance. Points represent nine patients. Boxes show the median and interquartile range, and whiskers extend to 1.5 times the interquartile range.\par
\endgroup

\section*{Discussion}

Our study establishes a framework in which open-ended biological discovery can be operationalized as an auditable and systematic process without being instructed by a prespecified biological question. PROSPECTor enables candidate signals to emerge from data-driven exploration and to be progressively refined through layers of evidence, including reproducibility, molecular interpretation and robustness assessment. Across our analyses, search strategies and heterogeneous single-cell representation spaces generated independently supported hypotheses, including candidates with transferable molecular and cross-cohort evidence that warrant future experimental investigation.

Interestingly, representation heterogeneity reshapes the observable landscape of single-cell discovery. Each scFM embedding may encode a distinct inductive bias over the same molecular measurements, altering which neighbourhood structures, functional axes and rare cellular states become accessible to search. This view extends the use of scFM representations beyond predictive downstream applications, suggesting that pretrained models can serve as complementary discovery substrates in which the same molecular measurements are reorganized into alternative latent spaces. Yet representation-dependent signals are not interpreted directly from the embedding itself, since biological relevance is still determined through evidence acquired after discovery. Our case studies illustrate this by translating representation-specific or detector-specific structures into testable gene programmes that could be evaluated across independent cohorts, intervention contexts and data modalities.

Several limitations remain. First, independent-cohort evaluation can support the generalizability and phenotype association of a candidate hypothesis, yet biological-unit-level effect estimates can remain imprecise, and such evaluation cannot by itself validate the putative biological function or causal mechanism. Orthogonal bulk, spatial and intervention-cohort analyses can extend this evidence profile, but each retains its limits. Second, functional interpretation remains dependent on current biological knowledge, and programmes with uncharacterized functions, regulatory mechanisms or cellular roles may remain difficult to interpret. Third, expanding the search space increases the need for expert oversight. Investigators remain essential for evaluating cohort suitability, interpreting evidence across modalities, selecting functional assays and determining which candidates justify experimental investment.

Future developments should strengthen feedback loops between computational discovery, biological interpretation and experimental validation. Modularity of PROSPECTor enables integration of emerging single-cell representation spaces, alternative data modalities, and experimental feedback. More broadly, this work suggests a reversal of the conventional question-driven order of analysis and a shift in how existing biological measurements can be used: beyond serving as retrospective archives generated to address predefined questions, they can become prospective resources from which new research questions emerge.

\section*{Methods}

\subsection*{Overview of the PROSPECTor design}

PROSPECTor analysis comprised candidate discovery, hypothesis fixation, independent validation and post-validation reporting. A candidate became eligible for independent evaluation only after its numerical definition had been recorded, including its genes or representation loadings, signed weights and orientation, cellular scope, normalization, threshold and aggregation rule where applicable. For projectable gene programmes, the fixed genes and signed weights defined a cell-level candidate score after the registered expression normalization. These scores were subsequently aggregated within samples and biological units according to the recorded rule. Independent data were used only to apply this fixed definition. After all external evidence had been assembled, the final reporting step could update the ranking of the candidate set.

The independent biological unit was the donor, patient or animal, as appropriate to the study design. Cell-level measurements were first summarized within samples, and samples were then weighted equally within each biological unit. Detection was performed at the cell level, whereas recurrence and validation were assessed at the biological-unit level. For paired human data, replicate samples with the same tissue role were averaged within patient before tumour-minus-adjacent differences were calculated. Patients lacking either member of a pair were excluded from paired endpoints.

\subsection*{Data intake, information isolation, and preprocessing}

The workflow accepted H5AD, 10x HDF5, Matrix Market and 10x directory formats, plain or gzip-compressed CSV or TSV matrices, and GEO RAW tar archives. Each run specified the discovery files, organism, tissue and fields identifying cells, samples and donors. A separate manifest specified any independent single-cell cohorts requested for external evaluation.

At data intake, the workflow determined whether the expression matrix contained integer counts, identified the cell and gene axes, standardized gene identifiers and verified that every cell mapped to one sample and every sample to one biological unit. When multiple matrices were supplied, they were merged over the union of genes while preserving study, sample and donor boundaries.

During discovery, the workflow received the expression matrix together with anonymous sample, donor and study-partition identifiers. The workflow did not receive condition, treatment or tumour-versus-adjacent labels, study accession, paper title, source-paper conclusions, independent-study data, evaluation outputs or literature-retrieval results. The mapping between source identifiers and anonymous identifiers was kept outside the discovery workspace. For paired designs, the workflow could retain information that two anonymous samples belonged to the same participant so that within-person contrasts could be constructed, while the phenotype labels of the paired samples remained hidden. This information boundary was enforced at the workspace level. The discovery report and projection specification were finalized before the evaluation workspace made phenotype information available. Phenotype labels could subsequently be used to describe or test a fixed candidate.

For conventional expression analyses, OmicVerse\textsuperscript{54} normalized counts to 10,000 per cell, applied a log(1+x) transformation, selected 2,000 highly variable genes using the Seurat dispersion procedure and performed PCA, retaining the first 50 components as the expression representation. The selected genes, centring, scaling and PCA loadings were then applied consistently across all discovery-study partitions. Raw integer counts were retained for count-dependent analyses.

Quality-control records included raw library size, detected-gene count, mitochondrial fraction and the number of retained cells for each sample and biological unit. Thresholds were either specified in the run request or taken from the registered procedure and were stored in the run record. Existing cell annotations and pretrained embeddings could be supplied when their cell identities and ordering matched the retained count matrix.

For cell annotations, broad cell lineages were assigned before finer cell types. Cell-type-restricted analyses used only confidently assigned labels. Human annotation used fixed Enrichr exports of CellMarker 2024, PanglaoDB Augmented 2021 and Tabula Sapiens marker libraries.\textsuperscript{55–58} Mouse annotation used the corresponding mouse terms from CellMarker 2024 and PanglaoDB Augmented 2021, together with Tabula Muris.\textsuperscript{59} Mouse gene symbols were mapped to official MGI symbols using a pinned MGI marker list. Documented synonyms were allowed, but human-to-mouse orthologue substitution was not used.\textsuperscript{60}

\subsection*{Single-cell foundation-model representations}

The full scFM representation spaces contained cell embeddings extracted from scGPT, Geneformer, scFoundation, UCE, Tahoe-X1 and STATE.\textsuperscript{6,19–23} All six were extracted from the same cells with frozen pretrained checkpoints through the CellEncoder inference adapters (\href{https://github.com/Open-BioAI/CellEncoder/}{\url{https://github.com/Open-BioAI/CellEncoder/}}). Gene identifiers were converted to the namespace required by each model before inference. scGPT, UCE, scFoundation and STATE used gene symbols, whereas Geneformer and Tahoe-X1 used Ensembl identifiers.

The resulting embeddings had 512 dimensions for scGPT, 1,152 for Geneformer-V2-316M, 3,072 for scFoundation, 1,280 for the four-layer UCE model, 512 for Tahoe-X1-70M and 2,058 for STATE. UCE used a token length of 1,536, a batch size of 64, a sampling size of 64 and seed 0. Tahoe-X1 used a batch size of 64, and STATE used a batch size of 128. For mouse single-cell data, only UCE embeddings were extracted. Each scFM representation was searched independently in its native dimensionality and after a separately fitted 32-component PCA reduction.

\subsection*{Detector parameters}

\begingroup
\small
\sloppy
\setlength{\tabcolsep}{3pt}
\renewcommand{\arraystretch}{1.12}
\begin{longtable}{@{}>{\RaggedRight\arraybackslash}p{0.228\linewidth}>{\RaggedRight\arraybackslash}p{0.228\linewidth}>{\RaggedRight\arraybackslash}p{0.228\linewidth}>{\RaggedRight\arraybackslash}p{0.228\linewidth}@{}}
\toprule
\textbf{Operator} & \textbf{Numerical definition} & \textbf{Stability and eligibility} & \textbf{Exported candidate} \\
\midrule
\endfirsthead
\toprule
\textbf{Operator} & \textbf{Numerical definition} & \textbf{Stability and eligibility} & \textbf{Exported candidate} \\
\midrule
\endhead
Stable population & 30-nearest-neighbour graph, Leiden resolution 0.6,\textsuperscript{61} at most 12 clusters & Seeds 17 and 29, two 80\% cell resamples, minimum cell fraction 0.03, minimum agreement 0.35 & Fixed cell membership and donor burden \\
Sparse direction & Four components, elastic-net mixing 0.7, ridge 0.01, maximum 900 LARS iterations, tolerance 10\textsuperscript{-7} & Upper and lower 5\% tails, resample score correlation ≥0.20, loading-support Jaccard ≥0.10 & Continuous cell score, tails and representation loadings \\
Rare state & Euclidean 20-nearest-neighbour distance, upper 5\% rarity tail  & At least 24 cells and four donors, no donor contributing >50\% of the state, resample score correlation ≥0.20, membership Jaccard ≥0.10 & Fixed rarity score, membership and donor burden \\
Expression programme & cNMF rank 3, 500 HVGs, density threshold 2.0, two registered seeds & Upper 10\% usage, at least 24 cells, four donors and five non-technical genes & Continuous usage and top 30 fixed genes \\
\bottomrule
\end{longtable}
\endgroup

\subsection*{Candidate refinement and evidence acquisition}

Each detector output received a fixed candidate identifier, eligible-cell set and numerical definition. Follow-up analyses were determined by candidate type and the evidence already available. These included cell-type conditioning, nuisance balancing, removal of components predictable from conventional expression, repeated-sample contrasts, recurrent-subgroup analysis, cross-representation comparison and molecular decoding into a signed gene programme. Failure of detector stability, biological-unit coverage or required multi-study support closed the corresponding branch. An analysis that could not be performed because the required measurements were unavailable was recorded as unresolved.

Refinement created a new candidate only when an analysis changed the numerical definition or cellular scope of the parent signal. Each successor was evaluated independently and had to satisfy its own stability, biological-unit coverage and downstream evidence requirements. Parent evidence was retained as provenance but was not inherited as support. Cell-type conditioning required both candidate arms to contain at least two cells per arm in at least four biological units, with no biological unit contributing more than 50\% of retained cells. At most six cell-type subsets were retained, ordered by biological-unit coverage and cell count. Nuisance balancing matched candidate arms within biological unit and across five sequencing-depth quantile bins. It required at least two cells per arm in four retained units and retention of at least 20\% of eligible cells, followed by 100 within-stratum permutations. Repeated-sample programmes required exactly two anonymous samples from each of at least six donors. They were retained when the minimum leave-one-donor programme cosine was at least 0.5, the median held-out pair-projection cosine was at least 0.2 and the maximum absolute within-donor correlation with mean library size or retained-cell count was below 0.7. Recurrent subgroups required at least six biological units and three units per subgroup, standardized separation of at least 1.5, leave-one-unit assignment agreement of at least 0.75 and bootstrap agreement of at least 0.80. When several studies were present, no single study could contribute more than 85\% of either subgroup.

Candidate similarity was used separately for cross-representation comparison and final de-duplication. Stable populations and rare states were compared by membership Jaccard, whereas continuous directions were compared by absolute Spearman correlation. Frozen hypotheses were considered corresponding when target-cell-context Jaccard was at least 0.50 and the absolute cosine similarity between their signed gene programmes was at least 0.25. Final candidate statements were grouped only when every pair had a stop-word-filtered token Jaccard of at least 0.60. The representative was selected using a fixed priority order based on evidence status, representation support, successor generation, measured evidence and available falsification tests. Refinement continued until every applicable analysis was completed, rejected or marked unavailable.

\subsection*{Molecular interpretation}

Molecular decoding related each fixed candidate to measured gene expression. Cells assigned to a stable or rare population were compared with eligible non-member cells. Continuous axes were evaluated using both their fixed upper and lower tails and association with the continuous score. For cNMF candidates, programme usage and the orientation of gene loadings were retained unchanged.

Gene-level contrasts were first calculated within biological units and then combined across units. For each gene, the output recorded the mean effect, number of informative donors, fraction of donors supporting the same direction and Benjamini–Hochberg-adjusted statistical evidence. The same molecular contrast was repeated within sufficiently represented cell types to distinguish a localized transcriptional state from a change in cell abundance.

The complete signed gene ranking was used for gene-set enrichment analysis. Human analyses used Enrichr exports of KEGG 2021 Human and WikiPathways 2024 Human, and mouse analyses used the registered KEGG Mouse 2019 and WikiPathways Mouse 2024 sets after conversion to official MGI symbols.\textsuperscript{55,62,63} OmicVerse performed permutation-based enrichment using 1,000 permutations, seed 17, gene-set sizes between 15 and 500 genes and a requirement of at least three leading-edge genes. Pathways were reported at a Benjamini–Hochberg false-discovery rate of 0.05 together with the genes contributing to each term. Mitochondrial, ribosomal and other technical terms were retained as diagnostic annotations.

\subsection*{Candidate projection into external datasets and score computation}

Before an external cohort was accessed, each projectable gene programme was frozen with respect to its selected genes, signed weights, intercept where applicable, normalization, direction, eligible cellular scope and aggregation rule. For each eligible cell (c), expression counts were normalized to 10,000 per cell and transformed as log(1+x). The candidate score was then calculated from the fixed programme as

\begin{equation*}
z_c=b+\sum_{g\in G_{\mathrm{shared}}} w_g x_{cg},
\end{equation*}

where $w_g$ is the frozen signed weight of gene $g$, $x_{cg}$ is its normalized expression in cell $c$, $b$ is the fixed intercept when defined and $G_{\mathrm{shared}}$ contains programme genes available in the external dataset. Missing genes were not imputed, and the remaining weights were not refitted or reoriented. For each candidate–cohort pair, the fraction of selected genes present and the fraction of absolute programme weight retained were recorded. Projection proceeded only when the candidate-specific coverage criterion and the required organism mapping and cellular scope were satisfied. Other candidate–cohort pairs were recorded as incompatible or not projectable and remained in the exported evidence matrix.

Cell-level candidate scores were averaged within each sample over the cells specified by the frozen cellular scope. Whole-sample candidates used all eligible retained cells, whereas cell-type-restricted candidates used only cells assigned to the declared population. When several samples from the same biological unit were available, sample-level scores were weighted equally to obtain one donor-, patient- or animal-level score. Binary phenotype comparisons used these biological-unit scores to calculate group effects. For paired analyses, replicate samples with the same tumour-versus-adjacent label were first averaged within patient and the paired score difference was then calculated. Continuous endpoints were likewise evaluated at the biological-unit level. Quality-control and potential-confounder summaries, including retained cell counts, biological-unit coverage, library size, mitochondrial fraction and declared composition covariates, were reported alongside the candidate result.

\subsection*{Supported candidates@4}

The direction used for held-out evaluation was fixed from the discovery studies. Let $g_j$ denote the signed Hedges effect for a candidate in discovery study $j$, and let $\bar g$ denote the equal-study mean of these effects. The discovery orientation was defined as $s=+1$ when $\bar g \geq 0$ and $s=-1$ otherwise. Candidate scores in the held-out study were oriented by $s$ before the effect was calculated, so that a positive held-out effect indicates agreement with the direction observed across the discovery studies. The same orientation was used for the original and marker-removed programmes. Held-out phenotype labels could not alter this orientation, the selected genes or their weights.

Reference human and mouse cell-marker libraries were used to determine whether a transferred effect was dominated by a familiar broad cell-identity programme. For candidate gene set $G$, signed weights $w_g$ and marker term $M$, weighted overlap is defined as

\begin{equation*}
O_M=
\frac{\sum_{g\in G\cap M}|w_g|}
{\sum_{g\in G}|w_g|}.
\end{equation*}

The reference marker term with the largest weighted overlap was selected, and genes in $G \cap M$ were removed. Candidate scores were then recalculated using the remaining frozen genes and weights with the same normalization, orientation and aggregation rule. No coefficients were refitted. This second projection tested whether the held-out effect persisted after removal of the candidate genes most strongly overlapping a reference cell-marker programme.

For an unpaired held-out study, the oriented Hedges effect was calculated from the biological-unit scores as the difference between phenotype-group means divided by the pooled standard deviation, with the Hedges small-sample correction.\textsuperscript{45} For paired cancer studies, tumour and adjacent scores were first averaged within patient, and the standardized effect was calculated from the resulting patient-level tumour-minus-adjacent differences with the corresponding small-sample correction. Positive effects indicate agreement with the discovery-defined direction in both settings.

Supported candidates@4 was defined as the number of candidates among the first four hypotheses in the frozen pre-validation ranking for which both the original and marker-removed oriented Hedges effects were at least 0.5. When a method returned fewer than four candidates, its observed count and corresponding lower attainable maximum were retained. A Hedges effect of 0.5 was specified as the primary threshold. Sensitivity analyses repeated the calculation at thresholds of 0.2, 0.3 and 0.8. Fold-level counts were first averaged within each disease setting, after which the four setting-level means received equal weight. The additional evidence measurements described below and source-study rediscovery were evaluated separately and did not contribute to Supported candidates@4.

\subsection*{Additional evidence measures}

Cross-source consistency measured whether discovery studies supported a common direction. With signed effects $g_j$, consistency was calculated as

\begin{equation*}
C=\frac{|\bar{g}|}{\frac{1}{J}\sum_j |g_j|},
\end{equation*}

with zero assigned when the denominator was zero. The value approaches one when source-study effects share a direction and approaches zero when effects of similar magnitude cancel. It was interpreted together with effect magnitude.

Cellular localization was evaluated without phenotype labels. For donor (d), the localization contrast was

\begin{equation*}
l_d=\bar{z}_{d,T}-\bar{z}_{d,\neg T},
\end{equation*}

where $T$ denotes the declared cell population and the complement contains all other confidently annotated cells from the same donor. Let $p$ denote the larger of the fractions of positive and negative $l_d$, and let

\begin{equation*}
e=\frac{|\mathrm{mean}(l_d)|}{\mathrm{sd}(l_d)}.
\end{equation*}

The localization score was the arithmetic mean of $\max(0,2p-1)$ and $\tanh(e/2)$. High values therefore required both consistency across donors and separation between the declared cell population and its within-donor complement. This quantity describes where a programme is concentrated and is distinct from the abundance of the target cell population.

Phenotype association within the declared population used each donor's mean fixed candidate score among target cells. Population-abundance association used the fraction of confidently annotated cells assigned to the same target cell population. Each endpoint was summarized using direction-invariant AUROC, prevalence-corrected AUPRC, leave-one-donor sign stability and $\tanh(|g|/2)$, where $g$ denotes the absolute Hedges effect, with their arithmetic mean forming the endpoint score. For paired gastric studies, within-patient differences were calculated before these measurements were obtained. These secondary scores were direction-invariant because they assess whether the declared cellular interpretation contains phenotype-associated information.

Technical robustness measured the fraction of the whole-sample phenotype effect retained after separately removing associations with mean log-transformed library size, mean log-transformed detected-gene count and mean mitochondrial fraction. For each technical covariate, donor candidate scores were regressed on an intercept and the standardized covariate without including phenotype in the regression. When the case–control contrast in the residuals reversed direction, retention was set to zero. Otherwise, retention was calculated as

\begin{equation*}
\min\left(1,\frac{|\Delta_{\mathrm{residual}}|}{|\Delta_{\mathrm{raw}}|}\right).
\end{equation*}

Technical robustness was the mean of the available retention values. Composition robustness applied the same procedure using the fraction of cells assigned to the declared population. Alternative-explanation resistance was defined as the smaller of mean technical retention and composition retention when both were available, preventing strong performance on one class of explanation from masking sensitivity to the other.

\subsection*{Source-study rediscovery}

Rediscovery targets were registered from each source paper before method outputs were scored and were decomposed into four layers comprising cell type or state, signed molecular programme, study direction or contrast, and biological process or inter-compartment relation. A deterministic procedure matched each registered finding one-to-one to at most one of the corresponding method's top-four structured candidates, preventing a single broad claim from receiving credit for multiple findings.

Cell types and states were compared using stop-word-filtered token F1. Signed molecular programmes used exact, direction-preserving gene-symbol recall together with pathway-token F1, while direction or contrast and process or relation were compared using their respective structured fields. Applicable components, findings and source studies received equal weight. Language-model semantic matching and biological-expert review were retained as separate audits and did not contribute to the six-study rediscovery value shown in Fig. 2d.

\subsection*{Evaluation scenarios and cohort selection}

Cohorts were selected according to study design, public accessibility and availability of biological-unit metadata (Supplementary Table 1). Before common evaluation preparation, we reproduced the quality-controlled analysis object used or explicitly released by each source paper where available.

The COVID-19 setting comprised six CELLxGENE blood or PBMC collections with primary single-cell data, donor identities and case–control metadata.\textsuperscript{25–31} Collection UUIDs, dataset UUIDs and donor counts are reported in Supplementary Table 1. Eligibility for primary held-out evaluation required at least three cases and three controls. D02–D06 met this criterion. D01 contained 14 cases and one control and was retained as a sensitivity fold and as a discovery source, but it was excluded from the primary setting average. The IPF setting comprised GSE122960, GSE128033, GSE135893 and GSE136831. Each study provided public raw counts, auditable donor identities and an IPF–control lung comparison.\textsuperscript{32–35} The mouse-colon setting combined the public studies GSE264408 and GSE172261 with an independent Control, AOM–DSS and AOM–DSS + BAPN cohort previously generated within the study team.\textsuperscript{36–38} Together, these datasets span related inflammatory and tissue-remodelling biology while differing in disease model, intervention and sampling design. The gastric-cancer setting used GSE150290, GSE183904 and GSE270680.\textsuperscript{39–41} The studies contained 23, 9 and 21 pairs, respectively.

The human-cancer rediscovery setting comprised GSE167297, GSE150290, GSE234129, GSE183904, GSE270680 and GSE236696 (R01–R06).\textsuperscript{39–44} These studies were selected to span different tissues and source-paper questions with findings that could be registered before analysis. Each study was evaluated as an independent rediscovery task. Supplementary Table 1 records the operational contrast, paired-patient count and summarized source-paper target for each task.

\subsection*{Reference workflows}

The two reference workflows used GPT-5.6-Sol with medium reasoning through isolated provider interfaces. Each workflow received the same expression input and was required to return the common structured discovery report and projection specification. The projection contract fixed candidate order, normalization, intercept, signed weights and cell-to-sample-to-unit aggregation. Workflows could return four distinct candidates, a fifth candidate when justified, or fewer candidates when additional outputs would represent unsupported duplication.

CellVoyager used three requested analyses and at most six iterative notebook cycles. Self-critique was enabled, while visual-language, retrieval and documentation extensions were disabled. Biomni used its official A1 agent with the same model and reasoning effort and without the optional 11-GB retrieval data lake.\textsuperscript{46}

\subsection*{Pre-validation candidate ranking for evaluation}

For cross-workflow evaluation, once the deterministic PROSPECTor search was complete, all candidates were presented to a single GPT-5.6-Sol call with medium reasoning, which selected four candidates (or all available candidates when fewer than four were available) for computation of the evaluation metric, Supported candidates@4. Candidate selection prioritized statistical qualification, cellular localization, donor recurrence, stability, alternative-explanation tests, a fixed validation readout and non-redundancy.

\subsection*{External-cohort validation and reporting}

The full workflow performed external validation independently of the evaluation scorer described above. Every candidate with a complete and permitted projection definition was applied to every compatible single-cell cohort in the user-supplied manifest. After all external projections, coverage checks and quality-control summaries were complete, a deterministic procedure assembled candidate-level evidence for reporting.

Language-model reranking was optional. A single GPT-5.6-Sol call with medium reasoning was made only when \nolinkurl{use_llm_reranking: true} was explicitly specified in \nolinkurl{run.json} and a model-provider configuration was supplied. The model received the complete candidate set together with cohort status, principal validation measurements, gene coverage, measured confounders and the most discriminating genes, pathways and cellular localizations. When reranking was disabled or no provider configuration was supplied, the workflow retained the deterministic candidate ordering and completed external validation and report generation without a language-model call. Thus, no model API credentials were required for end-to-end execution.

The workflow wrote corresponding JSON, TSV and Markdown reports. The final user-facing report summarized the input datasets, donor counts, comparison groups, biological research areas represented by the first four candidates and the complete candidate ranking, followed by the original deterministic discovery report.

\subsection*{scFM representation analyses}

F04 was selected for the descriptive scFM representation analysis after the fold-level analysis in Fig. 2f showed that four of its six phenotype-associated candidates originated from scFM representations, the highest number among the four IPF folds. For Fig. 3a, the same 10,000 F04 cells were used for every representation. To enable comparable visualization and diagnostic analyses across embeddings of different dimensionality, each cell vector was L2-normalized and representations with more than 50 dimensions were reduced to 50 components by truncated singular-value decomposition before UMAP and the analyses in Fig. 3b,c.

Cross-donor cell identity was measured by assigning each cell its nearest neighbour from a different donor and recording whether the two cells shared the same broad annotation. For the cross-study disease analysis, each representation was aggregated within donor and broad cell type, and class-balanced logistic regression was fitted under leave-one-study-out evaluation, with held-out AUROC reported for each representation. These two measurements were repeated in 20 donor-stratified 80\% resamples. Donor mixing was defined as the fraction of the 15 nearest neighbours contributed by other donors. Technical encoding was measured using fivefold donor-grouped ridge regression to predict detected-gene count from each representation, with cross-validated R\textsuperscript{2} reported. Pairwise neighbourhood concordance was calculated as the mean cell-level Jaccard overlap between 30-nearest-neighbour sets for the same cells. Supplementary Fig. 2a repeated the coarse identity and disease analyses after aggregation of all 90,776 discovery cells by donor and broad cell type. Supplementary Fig. 2d used average-linkage clustering with one minus neighbourhood concordance as the distance. The UCE example used the detector's continuous rarity score and fixed upper-5\% membership for discovery visualization. Independent-study evaluation used only the molecularly decoded 25-gene programme. Representation-specific counts of final candidates and candidates meeting the independent phenotype-association criterion of \emph{q} ≤ 0.05 were obtained from the candidate and validation records. Supplementary Fig. 2b,c used the same fixed candidates to assess recurrence in different representations and the component retained after comparison with conventional HVG-PCA.

\subsection*{Mouse-colon case study}

The mouse-colon analysis used GSE264408 and GSE172261, together comprising 16 discovery mice.\textsuperscript{36,37} Candidate \#1 was a stable fibroblast-associated state anchored by Pcolce2, Rbp1 and Serpinf1, whereas candidate \#2 was a continuous sparse programme anchored by Tgfbi, Igf2 and Bmp3. For the discovery displays, the raw detector readout was summarized per mouse within each study: cluster burden for candidate \#1 and median sparse-direction score for candidate \#2. Both signals were measurable in all 16 discovery mice.

The held-out validation dataset, previously generated within the study team, comprised three mice in each group.\textsuperscript{38} Control versus AOM–DSS was the independent case–control comparison, while AOM–DSS + BAPN was reserved as a subsequent intervention context. All six candidates were evaluated in every Control and AOM–DSS mouse using unchanged normalization and weights. The reported effect was AOM–DSS minus Control Hedges \emph{g}, including the small-sample correction. Its approximate 95\% confidence interval was the estimate plus or minus 1.96 times the normal-approximation standard error $\sqrt{(n_1+n_0)/(n_1n_0)+g^2/[2(n_1+n_0-2)]}$.\textsuperscript{45}

The whole-sample score was the mean fixed cell-level candidate score across all retained cells from each mouse. Cell-type-restricted analysis used a fibroblast annotation containing 3,066 cells across all nine mice. Candidate \#1 and \#2 scores were recomputed by averaging the same fixed cell-level candidate score within fibroblasts, and fibroblast abundance was the percentage of retained, confidently annotated cells assigned to fibroblasts. Whole-sample score, fibroblast abundance, fibroblast-restricted candidate \#1 and fibroblast-restricted candidate \#2 were kept as separate readouts as they distinguish a tissue-wide change, a compositional change and two cell-intrinsic programme changes. Candidate \#1 and \#2 retained their original score scales, so absolute values were compared within candidates rather than between panels.

Candidate-score movement under BAPN was oriented towards Control and defined as $R=(\bar z_{\mathrm{AOM–DSS+BAPN}}-\bar z_{\mathrm{AOM–DSS}})/(\bar z_{\mathrm{Control}}-\bar z_{\mathrm{AOM–DSS}})$. A value of zero denotes the AOM–DSS mean, one denotes the Control mean and values outside this range indicate movement beyond either reference mean. The formula was applied unchanged to every available PROSPECTor and reference-workflow candidate signal. Because each arm contained three unpaired mice, these intervention results were presented as descriptive individual-mouse measurements.

\subsection*{Gastric-cancer T-cell case study}

The candidate signal examined in this case study was decoded to give a fixed 24-gene T-cell programme. Its raw coefficients placed \emph{TTC3}, \emph{ATP1B1} and \emph{ITM2B} at one end and \emph{PSMB8}, \emph{PARP9}, \emph{SAMD9L} and \emph{GBP1} at the antigen-processing and interferon-stress end. For all case-study analyses, the candidate score was oriented once as the negative of the raw weighted sum so that higher values indicate the interferon-stress pole. This fixed orientation was used for all subsequent single-cell, bulk and spatial projections.

The primary T-lineage gate required detected \emph{PTPRC}, detection of at least two genes among \emph{CD3D}, \emph{CD3E}, \emph{TRAC}, \emph{CD247}, \emph{LCK}, \emph{TRBC1} and \emph{TRBC2}, and at least three total counts across these T-cell genes. The mean per-gene T-marker count was also required to be no lower than the largest corresponding mean across fixed B-cell, myeloid, epithelial and stromal marker modules. The 24 fixed programme coefficients were applied after counts-per-10,000 normalization and $\log(1+x)$ transformation. Candidate scores were averaged within T-lineage cells in each library, and patients were required to have at least 30 eligible T-lineage cells in both tissue roles.

The discovery analysis used paired T-cell pseudobulks from eight patients in GSE183904, while GSE270680 was specified in advance for independent validation.\textsuperscript{40,41} Genes, coefficients, score orientation and cellular scope were fixed before tumour and adjacent labels from GSE270680 were revealed for phenotype-association testing. The whole-sample projection averaged the fixed candidate score across all retained cells in each sample and included 21 paired patients. For the T-cell-restricted analysis shown, candidate scores were averaged within annotated T cells in each sample. Replicate samples with the same tissue role were then averaged within patient, and one tumour-minus-adjacent score difference was calculated for each of the 16 patients with at least 30 T cells in both tissues.

Beyond the prespecified GSE270680 holdout, additional T-cell-restricted projections included GSE150290, GSE206785, GSE167297 and GSE234129, with 23, 16, 4 and 3 evaluable patients, respectively.\textsuperscript{39,42,43,51} Each patient contributed one paired tissue difference calculated using the same fixed programme and score orientation. GSE167297 and GSE234129 contained only four and three eligible tumour–normal pairs, respectively. Their patient-level differences, unadjusted means and direction counts were therefore reported descriptively.

The GSE150290 projection used 47 raw droplet matrices from patients Pat01–Pat24, with Pat21 lacking a tumour library. Barcodes were retained when they contained 500–20,000 unique molecular identifiers, at least 200 detected genes and a combined mitochondrial and haemoglobin fraction below 10\%. All 23 complete patient pairs met the fixed T-lineage coverage requirement. The full T-cell compartment analysed here was absent from the quality-controlled non-immune object supplied to the evaluation run. GSE150290 was therefore treated as an additional cellular-compartment projection rather than a new independent study. GSE206785 provided a processed expression matrix from 24 treatment-naive patients with matched tumour and adjacent normal gastric tissue.\textsuperscript{51} The distributed values were consistent with log-transformed counts and were inverted before application of the fixed counts-per-10,000 and $\log(1+x)$ normalization procedure. Sixteen patients met the fixed T-lineage coverage requirement. Sensitivity analyses applied the programme directly to the distributed log-scale values, weighted CD4 and CD8 compartments equally, adjusted paired differences for library size, detected-gene count and CD4 fraction, and omitted each of the 24 programme genes in turn.

For orthogonal bulk recurrence, the same mapped programme genes and fixed coefficients were applied to TCGA-STAD expression from 33 patients with matched primary tumour and adjacent tissue.\textsuperscript{52} Both samples from each patient were retained, and recurrence was summarized from patient-level tumour-minus-adjacent score differences in the direction fixed by the single-cell programme. No coefficient, threshold or score orientation was learned from the bulk data. This whole-tissue analysis evaluates recurrence of the fixed gene programme but does not localize the signal to T cells.

Spatial localization used ten eligible GSE251950 Visium sections from nine patients.\textsuperscript{53} A pan-T-cell score was defined using \emph{CD247}, \emph{CD3D}, \emph{CD3E}, \emph{IL7R}, \emph{LCK} and \emph{TRAC}, none of which occurs in the 24-gene candidate programme. Within each section, a spot was classified as T-rich when the T-cell module was the highest-scoring lineage module, its section-standardized score was at least zero and at least two pan-T-cell genes were detected. The primary spatial analysis compared the fixed candidate score between T-rich and other tissue spots, after which section-level estimates were averaged within patient. A sensitivity model included the continuous pan-T-cell score as a covariate and retained the T-rich coefficient only when the design matrix was full rank and passed the prespecified collinearity criterion. Patient-level directional consistency, bootstrap confidence intervals and exact sign-flip tests summarized the spatial effects.\textsuperscript{64} Because Visium spots contain multiple cells, this analysis supports localization to T-rich niches but does not establish T-cell-intrinsic expression. For descriptive visualization in Fig. 5d, we selected the eligible primary-tumour section with the greatest Jaccard overlap between independently defined T-rich spots and the upper within-section candidate-score quintile (Jaccard = 0.191; 44.2\% of T-rich spots occurred in that quintile). This section-selection rule did not enter the patient-level statistical analyses.

\subsection*{Statistical analysis}

All statistical analyses used the donor, patient or animal as the independent biological unit. For comparisons between two independent groups, effect size was reported as Hedges \emph{g}, which standardizes the group difference by the pooled standard deviation and includes a small-sample correction. For paired tumour–adjacent analyses, a score difference was first calculated within each patient and the paired effect was standardized across these patient-level differences. When the complete set of possible group assignments was computationally tractable, unpaired comparisons were additionally assessed by exact label permutation. Direction counts in paired analyses were assessed using exact two-sided sign tests. Fixed paired programmes were assessed by exact sign-flip tests of the patient-level differences. Tests were two-sided unless a candidate direction had been fixed before independent evaluation, while confidence intervals were two-sided throughout. Benjamini–Hochberg false-discovery-rate correction was applied within each registered family of gene-, pathway-, candidate- or endpoint-level tests. Cohorts containing four or fewer eligible biological units were reported descriptively and were not used for multivariable-adjusted inference.

In the independent three-versus-three mouse comparison, Hedges \emph{g} was accompanied by an approximate 95\% confidence interval and exact enumeration of all possible group-label assignments. Given the small number of animals, these results were interpreted descriptively together with the individual-mouse measurements. Confidence intervals and resampling procedures otherwise summarized uncertainty at the donor or patient level. Random seeds and analysis parameters were stored with each run record.

\subsection*{Software and reproducibility}

Analyses used \nolinkurl{prospector-discovery} version 0.1.0. Public runs can be initiated through the PROSPECTor CLI using a run-request file and an external-cohort manifest.

\section*{Data availability}

Public datasets used in the principal analyses are identified by accession or stable CELLxGENE UUID in Supplementary Table 1. These include the six COVID-19 collections D01–D06; GSE122960, GSE128033, GSE135893, GSE136831, GSE264408, GSE172261, GSE150290, GSE183904, GSE270680, GSE206785, GSE167297, GSE234129, GSE236696, GSE251950 and TCGA-STAD. The raw sequencing data for the mouse-colon cohort (ref. 38) have been deposited in the Genome Sequence Archive (GSA) under accession CRA045737.

\section*{Code availability}

PROSPECTor version 0.1.0 is available at \href{https://github.com/ningxuan-zhang/PROSPECTor/}{\url{https://github.com/ningxuan-zhang/PROSPECTor/}}. The repository includes example discovery requests, external-cohort manifests, documentation, and a public acceptance run.

\section*{Acknowledgements}

Foremost, N.Z. is deeply grateful to N.L. for introducing her to research, for years of unwavering support and intellectual guidance, and for countless conversations that shaped much of the conceptual foundation underlying this work. She thanks H.W. and S.X. for their generous discussions and thoughtful criticism throughout the development of the project. Finally, PROSPECTor took shape during an unusual period; N.Z. is grateful for the questions that emerged along the way, for the curiosity that persisted alongside them, for the directions they sometimes offered, and for the prospect of pursuing them wherever they may lead.

\section*{Supplementary tables}

\subsection*{Supplementary Table 1 | Dataset registry and selection}

\begingroup
\small
\sloppy
\setlength{\tabcolsep}{3pt}
\renewcommand{\arraystretch}{1.12}
\begin{longtable}{@{}>{\RaggedRight\arraybackslash}p{0.228\linewidth}>{\RaggedRight\arraybackslash}p{0.228\linewidth}>{\RaggedRight\arraybackslash}p{0.228\linewidth}>{\RaggedRight\arraybackslash}p{0.228\linewidth}@{}}
\toprule
\textbf{Setting} & \textbf{Studies} & \textbf{Use in this study} & \textbf{Notes} \\
\midrule
\endfirsthead
\toprule
\textbf{Setting} & \textbf{Studies} & \textbf{Use in this study} & \textbf{Notes} \\
\midrule
\endhead
COVID-19 PBMC & Six CELLxGENE collections, D01–D06; UUIDs below & Five primary study-level folds plus D01 sensitivity analysis & D01 has one control and is excluded from the primary average \\
Human IPF & GSE122960, GSE128033, GSE135893, GSE136831 & Four complete study-level folds & Fold F04 was later selected for the descriptive analysis in Fig. 3 because it contained the largest number of phenotype-associated scFM-derived candidates (Fig. 2f); evaluation results include all folds \\
Mouse colonic inflammation & GSE264408, GSE172261, CRA045737\textsuperscript{38} & Three complete study-level folds and the first case study & Independent study-team cohort; source ref. 38 \\
Gastric cancer & GSE150290, GSE183904, GSE270680; GSE206785, GSE167297, GSE234129 & Three paired study-level folds and the second case study & The evaluation used the source-paper quality-controlled object for GSE150290; its full raw T-cell compartment was subsequently analysed in the case study \\
Human-cancer rediscovery & R01–R06; accessions and registrations below & Six source-paper rediscovery runs & The selected studies may not be representative \\
\bottomrule
\end{longtable}
\endgroup

\clearpage
\begin{landscape}
\textbf{COVID-19 CELLxGENE source registry.} Case, control and excluded counts refer to retained biological units after application of the fixed case–control mapping; \nolinkurl{excluded} denotes donors not assigned to the registered binary contrast. The source manifest was frozen on 16 July 2026.\par
\vspace{0.6em}
\begingroup
\scriptsize
\sloppy
\setlength{\tabcolsep}{3pt}
\renewcommand{\arraystretch}{1.12}
\begin{longtable}{@{}>{\RaggedRight\arraybackslash}p{0.040\linewidth}>{\RaggedRight\arraybackslash}p{0.225\linewidth}>{\RaggedRight\arraybackslash}p{0.225\linewidth}>{\RaggedRight\arraybackslash}p{0.085\linewidth}>{\RaggedRight\arraybackslash}p{0.045\linewidth}>{\RaggedRight\arraybackslash}p{0.050\linewidth}>{\RaggedRight\arraybackslash}p{0.055\linewidth}>{\RaggedRight\arraybackslash}p{0.165\linewidth}>{\RaggedRight\arraybackslash}p{0.050\linewidth}@{}}
\toprule
\textbf{ID} & \textbf{CELLxGENE collection UUID} & \textbf{Dataset UUID} & \textbf{Retained primary cells} & \textbf{Cases} & \textbf{Controls} & \textbf{Excluded} & \textbf{Holdout status} & \textbf{Source ref.} \\
\midrule
\endfirsthead
\toprule
\textbf{ID} & \textbf{CELLxGENE collection UUID} & \textbf{Dataset UUID} & \textbf{Retained primary cells} & \textbf{Cases} & \textbf{Controls} & \textbf{Excluded} & \textbf{Holdout status} & \textbf{Source ref.} \\
\midrule
\endhead
D01 & \nolinkurl{0434a9d4-85fd-4554-b8e3-cf6c582bb2fa} & \nolinkurl{fa8605cf-f27e-44af-ac2a-476bee4410d3} & 59,506 & 14 & 1 & 0 & Sensitivity/source only; one group with $n=1$ & 26 \\
D02 & \nolinkurl{4f889ffc-d4bc-4748-905b-8eb9db47a2ed} & \nolinkurl{de2c780c-1747-40bd-9ccf-9588ec186cee} & 49,053 & 8 & 4 & 5 & Primary eligible & 27 \\
D03 & \nolinkurl{7d7cabfd-1d1f-40af-96b7-26a0825a306d} & \nolinkurl{01ad3cd7-3929-4654-84c0-6db05bd5fd59} & 521,099 & 54 & 11 & 15 & Primary eligible & 28 \\
D04 & \nolinkurl{8f126edf-5405-4731-8374-b5ce11f53e82} & \nolinkurl{ebc2e1ff-c8f9-466a-acf4-9d291afaf8b3} & 816,915 & 102 & 10 & 12 & Primary eligible & 29 \\
D05 & \nolinkurl{ddfad306-714d-4cc0-9985-d9072820c530} & \nolinkurl{c7775e88-49bf-4ba2-a03b-93f00447c958} & 632,209 & 86 & 29 & 5 & Primary eligible & 30 \\
D06 & \nolinkurl{ecb739c5-fe0d-4b48-81c6-217c4d64eec4} & \nolinkurl{242c6e7f-9016-4048-af70-d631f5eea188} & 189,902 & 5 & 9 & 0 & Primary eligible & 31 \\
\bottomrule
\end{longtable}
\endgroup
\end{landscape}
\clearpage

\clearpage
\begin{landscape}
\textbf{Human-cancer source-paper rediscovery registry.} The listed contrast is the operational paired endpoint used for rediscovery.\par
\vspace{0.6em}
\begingroup
\scriptsize
\sloppy
\setlength{\tabcolsep}{3pt}
\renewcommand{\arraystretch}{1.12}
\begin{longtable}{@{}>{\RaggedRight\arraybackslash}p{0.157\linewidth}>{\RaggedRight\arraybackslash}p{0.157\linewidth}>{\RaggedRight\arraybackslash}p{0.157\linewidth}>{\RaggedRight\arraybackslash}p{0.157\linewidth}>{\RaggedRight\arraybackslash}p{0.157\linewidth}>{\RaggedRight\arraybackslash}p{0.157\linewidth}@{}}
\toprule
\textbf{ID} & \textbf{Accession} & \textbf{Operational contrast} & \textbf{Complete patient-level pairs} & \textbf{Summarized source-paper target} & \textbf{Source ref.} \\
\midrule
\endfirsthead
\toprule
\textbf{ID} & \textbf{Accession} & \textbf{Operational contrast} & \textbf{Complete patient-level pairs} & \textbf{Summarized source-paper target} & \textbf{Source ref.} \\
\midrule
\endhead
R01 & GSE167297 & Deep versus superficial tumour & 5 & Invasion-depth gradient; CCL2 fibroblast–inflammatory endothelial relation & 42 \\
R02 & GSE150290 & Tumour versus adjacent & 23 & Lauren-type branch, precancer progression and epithelial lineage relation & 39 \\
R03 & GSE234129 & Primary tumour versus adjacent & 3 & SDC2 CAF state and metastatic continuity & 43 \\
R04 & GSE183904 & Tumour versus normal & 9 & Stage/histology-associated states and INHBA/FAP CAF programme & 40 \\
R05 & GSE270680 & Tumour versus normal & 21 & FDCSP/CCL19 stromal states, tertiary-lymphoid-structure context and stromal–lymphoid relation & 41 \\
R06 & GSE236696 & Mucinous colorectal tumour versus adjacent & 6 & MUC1 tumour programme, FGF7/THBS1 myofibroblasts and an immunosuppressive inter-compartment relation & 44 \\
\bottomrule
\end{longtable}
\endgroup
\end{landscape}
\clearpage

\subsection*{Supplementary Table 2 | End-to-end PROSPECTor input formats and outputs}

\begingroup
\small
\sloppy
\setlength{\tabcolsep}{3pt}
\renewcommand{\arraystretch}{1.12}
\begin{longtable}{@{}>{\RaggedRight\arraybackslash}p{0.455\linewidth}>{\RaggedRight\arraybackslash}p{0.455\linewidth}@{}}
\toprule
\textbf{Item} & \textbf{Supported values} \\
\midrule
\endfirsthead
\toprule
\textbf{Item} & \textbf{Supported values} \\
\midrule
\endhead
Discovery and external-validation datasets & H5AD; 10x HDF5; Matrix Market/10x bundle; CSV/TSV; GEO RAW tar \\
User-supplied identifiers & Cell/sample/donor fields; organism and tissue \\
Final output & Corresponding JSON, TSV and Markdown reports; complete post-validation ranking \\
\bottomrule
\end{longtable}
\endgroup

\subsection*{Supplementary Table 3 | Evaluation workflows and execution settings}

\begingroup
\small
\sloppy
\setlength{\tabcolsep}{3pt}
\renewcommand{\arraystretch}{1.12}
\begin{longtable}{@{}>{\RaggedRight\arraybackslash}p{0.228\linewidth}>{\RaggedRight\arraybackslash}p{0.228\linewidth}>{\RaggedRight\arraybackslash}p{0.228\linewidth}>{\RaggedRight\arraybackslash}p{0.228\linewidth}@{}}
\toprule
\textbf{Method} & \textbf{Model} & \textbf{Reasoning} & \textbf{Tools/retrieval} \\
\midrule
\endfirsthead
\toprule
\textbf{Method} & \textbf{Model} & \textbf{Reasoning} & \textbf{Tools/retrieval} \\
\midrule
\endhead
PROSPECTor & \nolinkurl{gpt-5.6-sol} & Medium & No model use during search; one outcome-hidden pre-validation ranking call and one post-validation synthesis call \\
CellVoyager & \nolinkurl{gpt-5.6-sol} & Medium & Released notebook/code workflow; retrieval and VLM disabled; 3 analyses; ≤6 notebook cycles; 200,000-cell working cap \\
Biomni & \nolinkurl{gpt-5.6-sol} & Medium & Native A1 tool/code loop; optional data lake disabled \\
\bottomrule
\end{longtable}
\endgroup

\section*{References}

\begin{enumerate}[leftmargin=*,itemsep=0.35em]

\item Heumos, L. et al. Best practices for single-cell analysis across modalities. \emph{Nat. Rev. Genet.} \textbf{24}, 550–572 (2023).

\item Luecken, M. D. \& Theis, F. J. Current best practices in single-cell RNA-seq analysis: a tutorial. \emph{Mol. Syst. Biol.} \textbf{15}, e8746 (2019).

\item Alber, S. et al. CellVoyager: AI CompBio agent generates new insights by autonomously analyzing biological data. \emph{Nat. Methods} \textbf{23}, 749–759 (2026).

\item Yanai, I. \& Lercher, M. A hypothesis is a liability. \emph{Genome Biol.} \textbf{21}, 231 (2020).

\item Yanai, I. \& Lercher, M. J. Openness guides discovery. \emph{Nat. Biotechnol.} \textbf{43}, 667–668 (2025).

\item Cui, H. et al. scGPT: toward building a foundation model for single-cell multi-omics using generative AI. \emph{Nat. Methods} \textbf{21}, 1470–1480 (2024).

\item Wu, J. et al. Biology-driven insights into the power of single-cell foundation models. \emph{Genome Biol.} \textbf{26}, 334 (2025).

\item He, F. et al. Harnessing the power of single-cell large language models with parameter-efficient fine-tuning using scPEFT. \emph{Nat. Mach. Intell.} \textbf{8}, 118–133 (2026).

\item Zhao, Y. et al. How different AI models understand cells differently. Preprint at bioRxiv \url{https://doi.org/10.64898/2026.01.29.702682} (2026).

\item Gold, M. P. et al. Scoring gene importance by interpreting single-cell foundation models. \emph{Nat. Biotechnol.} \url{https://doi.org/10.1038/s41587-026-03112-5} (2026).

\item Gottweis, J. et al. Accelerating scientific discovery with Co-Scientist. \emph{Nature} \textbf{655}, 487–496 (2026).

\item Ghareeb, A. E. et al. A multi-agent system for automating scientific discovery. \emph{Nature} \textbf{655}, 497–505 (2026).

\item Nouri, N., Artzi, R. \& Savova, V. An agentic AI framework for ingestion and standardization of single-cell RNA-seq data analysis. \emph{npj Artif. Intell.} \textbf{2}, 8 (2026).

\item Elemento, O. AI systems devise hypotheses and ways to test them. \emph{Nature} \textbf{655}, 313–314 (2026).

\item Zimmerman, K. D., Espeland, M. A. \& Langefeld, C. D. A practical solution to pseudoreplication bias in single-cell studies. \emph{Nat. Commun.} \textbf{12}, 738 (2021).

\item Squair, J. W. et al. Confronting false discoveries in single-cell differential expression. \emph{Nat. Commun.} \textbf{12}, 5692 (2021).

\item Luecken, M. D. et al. Benchmarking atlas-level data integration in single-cell genomics. \emph{Nat. Methods} \textbf{19}, 41–50 (2022).

\item Tran, H. T. N. et al. A benchmark of batch-effect correction methods for single-cell RNA sequencing data. \emph{Genome Biol.} \textbf{21}, 12 (2020).

\item Theodoris, C. V. et al. Transfer learning enables predictions in network biology. \emph{Nature} \textbf{618}, 616–624 (2023).

\item Hao, M. et al. Large-scale foundation model on single-cell transcriptomics. \emph{Nat. Methods} \textbf{21}, 1481–1491 (2024).

\item Rosen, Y. et al. Universal cell embedding provides a foundation model for cell biology. \emph{Nature} \textbf{656}, 183–191 (2026).

\item Gandhi, S. et al. Tahoe-X1: scaling perturbation-trained single-cell foundation models to 3 billion parameters. Preprint at bioRxiv \url{https://doi.org/10.1101/2025.10.23.683759} (2025).

\item Adduri, R. et al. Predicting cellular responses to perturbation across diverse contexts with State. Preprint at bioRxiv \url{https://doi.org/10.1101/2025.06.26.661135} (2025).

\item Kotliar, D. et al. Identifying gene expression programs of cell-type identity and cellular activity with single-cell RNA-seq. \emph{eLife} \textbf{8}, e43803 (2019).

\item CZI Cell Science Program et al. CZ CELLxGENE Discover: a single-cell data platform for scalable exploration, analysis and modeling of aggregated data. \emph{Nucleic Acids Res.} \textbf{53}, D886–D900 (2025).

\item Chan Zuckerberg Initiative Single-Cell COVID-19 Consortia et al. Single cell profiling of COVID-19 patients: an international data resource from multiple tissues. Preprint at medRxiv \url{https://doi.org/10.1101/2020.11.20.20227355} (2020).

\item Lee, J. S. et al. Immunophenotyping of COVID-19 and influenza highlights the role of type I interferons in development of severe COVID-19. \emph{Sci. Immunol.} \textbf{5}, eabd1554 (2020).

\item van der Wijst, M. G. P. et al. Type I interferon autoantibodies are associated with systemic immune alterations in patients with COVID-19. \emph{Sci. Transl. Med.} \textbf{13}, eabh2624 (2021).

\item COVID-19 Multi-omics Blood ATlas Consortium. A blood atlas of COVID-19 defines hallmarks of disease severity and specificity. \emph{Cell} \textbf{185}, 916–938.e58 (2022).

\item Stephenson, E. et al. Single-cell multi-omics analysis of the immune response in COVID-19. \emph{Nat. Med.} \textbf{27}, 904–916 (2021).

\item Ivanova, E. N. et al. mRNA COVID-19 vaccine elicits potent adaptive immune response without the acute inflammation of SARS-CoV-2 infection. \emph{iScience} \textbf{26}, 108572 (2023).

\item Reyfman, P. A. et al. Single-cell transcriptomic analysis of human lung provides insights into the pathobiology of pulmonary fibrosis. \emph{Am. J. Respir. Crit. Care Med.} \textbf{199}, 1517–1536 (2019).

\item Morse, C. et al. Proliferating SPP1/MERTK-expressing macrophages in idiopathic pulmonary fibrosis. \emph{Eur. Respir. J.} \textbf{54}, 1802441 (2019).

\item Habermann, A. C. et al. Single-cell RNA sequencing reveals profibrotic roles of distinct epithelial and mesenchymal lineages in pulmonary fibrosis. \emph{Sci. Adv.} \textbf{6}, eaba1972 (2020).

\item Adams, T. S. et al. Single-cell RNA-seq reveals ectopic and aberrant lung-resident cell populations in idiopathic pulmonary fibrosis. \emph{Sci. Adv.} \textbf{6}, eaba1983 (2020).

\item Hong, D. et al. Integrative analysis of single-cell RNA-seq and gut microbiome metabarcoding data elucidates macrophage dysfunction in mice with DSS-induced ulcerative colitis. \emph{Commun. Biol.} \textbf{7}, 731 (2024).

\item Jasso, G. J. et al. Colon stroma mediates an inflammation-driven fibroblastic response controlling matrix remodeling and healing. \emph{PLoS Biol.} \textbf{20}, e3001532 (2022).

\item Wang, Z. et al. Persistent post-inflammatory matrix stiffening defines a premalignant mechanical niche in ulcerative colitis-associated neoplasia. \emph{bioRxiv} \href{https://doi.org/10.64898/2026.06.17.733043}{doi: 10.64898/2026.06.17.733043} (2026).

\item Kim, J. et al. Single-cell analysis of gastric pre-cancerous and cancer lesions reveals cell lineage diversity and intratumoral heterogeneity. \emph{npj Precis. Oncol.} \textbf{6}, 9 (2022).

\item Kumar, V. et al. Single-cell atlas of lineage states, tumor microenvironment, and subtype-specific expression programs in gastric cancer. \emph{Cancer Discov.} \textbf{12}, 670–691 (2022).

\item Gao, S. et al. A spatially resolved atlas of gastric cancer characterises a lymphocyte-aggregated region. \emph{Nat. Commun.} \textbf{17}, 2059 (2026).

\item Jeong, H. Y. et al. Spatially distinct reprogramming of the tumor microenvironment based on tumor invasion in diffuse-type gastric cancers. \emph{Clin. Cancer Res.} \textbf{27}, 6529–6542 (2021).

\item Wang, R. et al. Evolution of immune and stromal cell states and ecotypes during gastric adenocarcinoma progression. \emph{Cancer Cell} \textbf{41}, 1407–1426.e9 (2023).

\item Zhou, H. et al. Integrating single-cell and spatial analysis reveals MUC1-mediated cellular crosstalk in mucinous colorectal adenocarcinoma. \emph{Clin. Transl. Med.} \textbf{14}, e1701 (2024).

\item Hedges, L. V. Distribution theory for Glass's estimator of effect size and related estimators. \emph{J. Educ. Stat.} \textbf{6}, 107–128 (1981).

\item Huang, K. et al. Autonomous biomedical research with an artificial intelligence agent. \emph{Science} \textbf{393}, eadz4351 (2026).

\item McInnes, L., Healy, J., Saul, N. \& Großberger, L. UMAP: Uniform Manifold Approximation and Projection. \emph{J. Open Source Softw.} \textbf{3}, 861 (2018).

\item Benjamini, Y. \& Hochberg, Y. Controlling the false discovery rate: a practical and powerful approach to multiple testing. \emph{J. R. Stat. Soc. B} \textbf{57}, 289–300 (1995).

\item Levi-Galibov, O. et al. Heat shock factor 1-dependent extracellular matrix remodeling mediates the transition from chronic intestinal inflammation to colon cancer. \emph{Nat. Commun.} \textbf{11}, 6245 (2020).

\item Kinchen, J. et al. Structural remodeling of the human colonic mesenchyme in inflammatory bowel disease. \emph{Cell} \textbf{175}, 372–386.e17 (2018).

\item Kang, B. et al. Parallel single-cell and bulk transcriptome analyses reveal key features of the gastric tumor microenvironment. \emph{Genome Biol.} \textbf{23}, 265 (2022).

\item Cancer Genome Atlas Research Network. Comprehensive molecular characterization of gastric adenocarcinoma. \emph{Nature} \textbf{513}, 202–209 (2014).

\item Lee, S. H. et al. Spatial dissection of tumour microenvironments in gastric cancers reveals the immunosuppressive crosstalk between CCL2+ fibroblasts and STAT3-activated macrophages. \emph{Gut} \textbf{74}, 714–727 (2025).

\item Zeng, Z. et al. OmicVerse: a framework for bridging and deepening insights across bulk and single-cell sequencing. \emph{Nat. Commun.} \textbf{15}, 5983 (2024).

\item Kuleshov, M. V. et al. Enrichr: a comprehensive gene set enrichment analysis web server 2016 update. \emph{Nucleic Acids Res.} \textbf{44}, W90–W97 (2016).

\item Hu, C. et al. CellMarker 2.0: an updated database of manually curated cell markers in human/mouse and web tools based on scRNA-seq data. \emph{Nucleic Acids Res.} \textbf{51}, D870–D876 (2023).

\item Franzén, O., Gan, L.-M. \& Björkegren, J. L. M. PanglaoDB: a web server for exploration of mouse and human single-cell RNA sequencing data. \emph{Database} \textbf{2019}, baz046 (2019).

\item Tabula Sapiens Consortium. The Tabula Sapiens: a multiple-organ, single-cell transcriptomic atlas of humans. \emph{Science} \textbf{376}, eabl4896 (2022).

\item Tabula Muris Consortium. Single-cell transcriptomics of 20 mouse organs creates a Tabula Muris. \emph{Nature} \textbf{562}, 367–372 (2018).

\item Baldarelli, R. M. et al. Mouse Genome Informatics: an integrated knowledgebase system for the laboratory mouse. \emph{Genetics} \textbf{227}, iyae031 (2024).

\item Traag, V. A., Waltman, L. \& van Eck, N. J. From Louvain to Leiden: guaranteeing well-connected communities. \emph{Sci. Rep.} \textbf{9}, 5233 (2019).

\item Kanehisa, M. \& Goto, S. KEGG: Kyoto Encyclopedia of Genes and Genomes. \emph{Nucleic Acids Res.} \textbf{28}, 27–30 (2000).

\item Martens, M. et al. WikiPathways: connecting communities. \emph{Nucleic Acids Res.} \textbf{49}, D613–D621 (2021).

\item Good, P. I. \emph{Permutation, Parametric and Bootstrap Tests of Hypotheses}. 3rd edn (Springer, 2005).

\end{enumerate}

\clearpage

\section*{Supplementary figures and legends}

\begin{center}
\includegraphics[width=\textwidth,height=0.72\textheight,keepaspectratio]{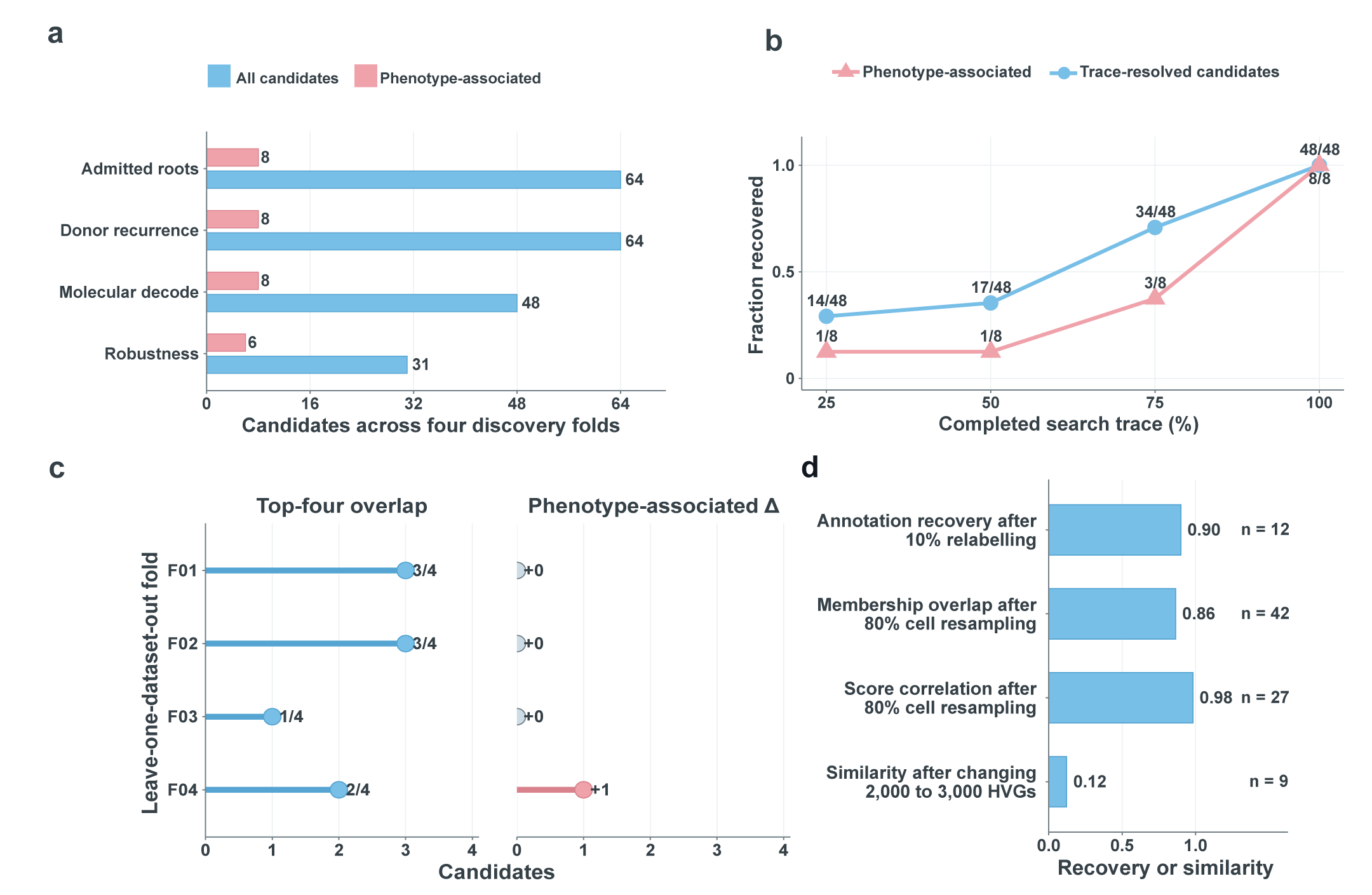}
\end{center}
\begingroup\small
\noindent\textbf{Supplementary Figure 1 | Additional evaluation of evidence acquisition, candidate prioritization and stability.} \textbf{a,} Candidate availability across successive evidence-acquisition stages. Blue bars show candidates with a recorded result at each stage, and coral overlays show the trace-resolved subset that subsequently met the independent-projection criterion of \emph{q} ≤ 0.05. Two additional candidates meeting this criterion lacked complete early-stage traces and are included only in the final total. \textbf{b,} Numbers of trace-resolved final candidates and phenotype-associated candidates recovered when each completed analysis trajectory was truncated after 25\%, 50\%, 75\% or 100\% of its recorded steps. \textbf{c,} Left, overlap between the deterministic evidence-ordered top four and the top four selected by a single outcome-hidden GPT-5.6-Sol call. Right, change in the number of phenotype-associated candidates after model selection. Positive values indicate more phenotype-associated candidates after selection. \textbf{d,} Candidate stability under 10\% annotation relabelling, two deterministic 80\% cell resamples and expansion of the HVG set from 2,000 to 3,000 genes. Points represent evaluable candidate records. Summary marks show the mean for annotation recovery and the median for the remaining analyses.\par
\endgroup

\clearpage
\begin{center}
\includegraphics[width=\textwidth,height=0.68\textheight,keepaspectratio]{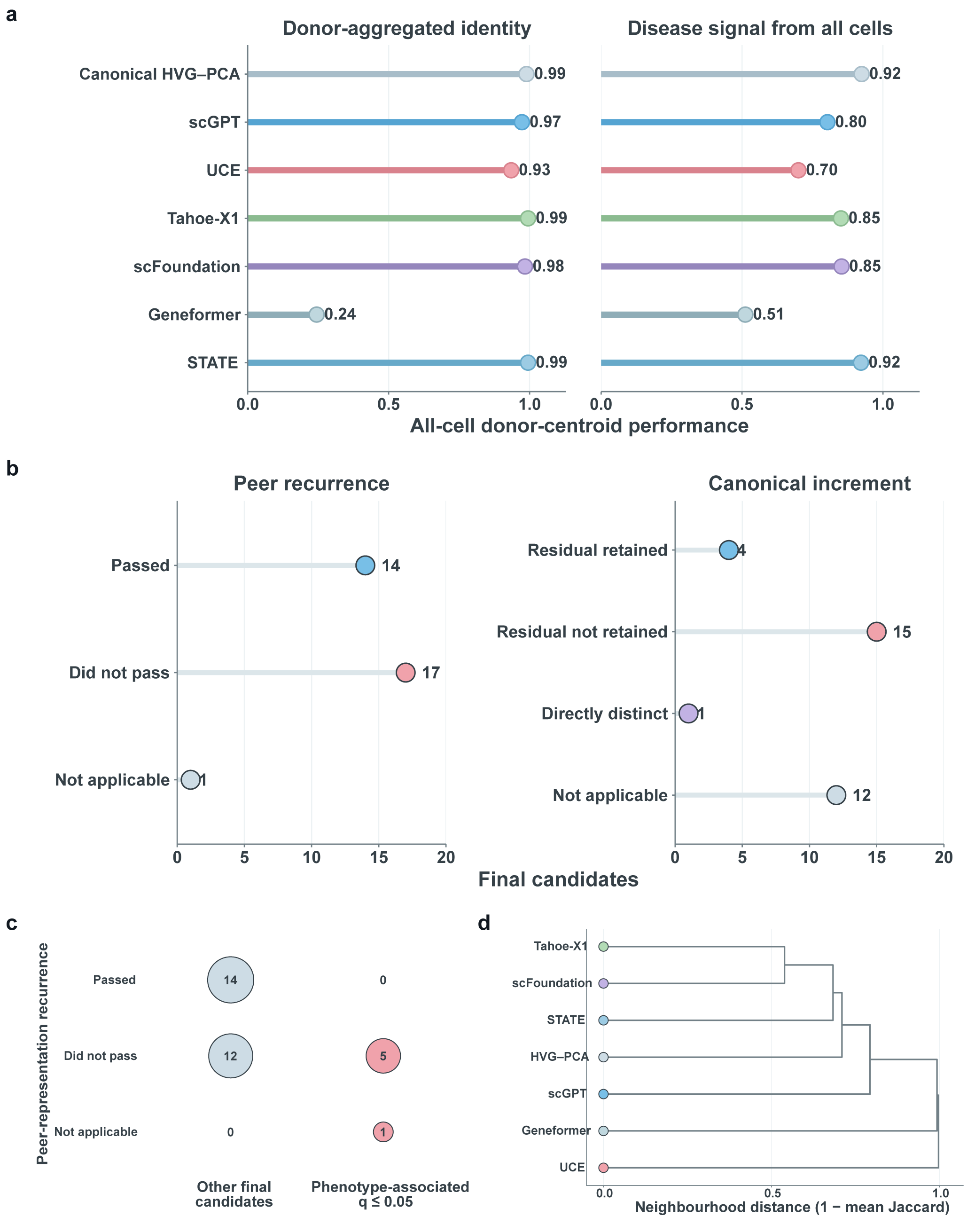}
\end{center}
\begingroup\small
\noindent\textbf{Supplementary Figure 2 | Cross-representation recurrence and clustering of scFM embeddings.} \textbf{a,} Broad cell identity and disease structure after aggregation of all 90,776 F04 discovery cells into 184 donor-by-broad-cell-type centroids. For donor-aggregated identity, each centroid was assigned the label of its nearest eligible centroid from another donor, and the plotted value is the fraction sharing the same broad cell identity. Disease signal is the mean AUROC from logistic classifiers trained on donor centroids within each broad cell type and evaluated with one study held out at a time. The two measurements use the same cells but assess different properties of a representation. Neither is used as a measure of candidate quality. \textbf{b,} Independent diagnostics for the 32 final candidates originating from scFM representations across the four IPF folds. Representation recurrence records whether the same donor-level endpoint was recovered in at least one other representation. Canonical increment records whether an evaluable component remained after comparison with conventional expression. Counts refer to candidate hypotheses. \textbf{c,} Cross-tabulation of cross-representation recurrence and independent phenotype association for the same 32 candidates. Circle area and printed values denote candidate counts. Independent phenotype association is defined by Benjamini–Hochberg-adjusted \emph{q} ≤ 0.05 after no-refit projection. Fourteen recurrence-positive candidates were not significant in the withheld study, whereas five recurrence-negative candidates and one candidate for which recurrence was not applicable were significant. \textbf{d,} Average-linkage clustering of the seven representations using one minus the mean cell-level Jaccard overlap between their 30-nearest-neighbour sets for the same 10,000 F04 cells. Branch length represents local-neighbourhood dissimilarity.\par
\endgroup

\end{document}